\documentclass[12pt,preprint]{aastex}
\usepackage{emulateapj5}
\usepackage{apjfonts}

\submitted{Accepted for Publication in the Astronomical Journal}

\shorttitle{First-Epoch Circular Polarization in MOJAVE}
\shortauthors{Homan and Lister}

\begin{document}

\title{MOJAVE: Monitoring of Jets in AGN with VLBA Experiments - II.
First-Epoch 15 GHz Circular Polarization Results}
\author{D. C. Homan}
\affil{Department of Physics and Astronomy, Denison University,
Granville, OH 43023; homand@denison.edu}
\and
\author{M. L. Lister}
\affil{Department of Physics, Purdue University, 525 Northwestern Avenue, 
West Lafayette, IN 47907; mlister@physics.purdue.edu}

\begin{abstract}
We report first-epoch circular polarization results for 133 active galactic nuclei in 
the MOJAVE program to monitor the structure and polarization of a flux limited sample of 
extra-galactic radio jets with the VLBA at 15 GHz.  We found strong circular polarization 
($\geq 0.3$\%) in approximately 15\% of our sample.  The circular polarization 
was usually associated with jet cores; however, we did find a few strong 
jet components to be circularly polarized.  The levels of
circular polarization were typically in the range of $0.3-0.5$\% of the 
local Stokes-$I$.   We found no strong correlations between fractional circular 
polarization of jet cores and source type, redshift, EGRET detections, linear polarization, 
or other observed parsec-scale jet properties.  There were differences between
the circular-to-linear polarization ratios of two nearby galaxies versus more distant
quasars and BL Lac objects.  We suggest this is because the more distant
sources either have (1) less depolarization of their linear polarization, and/or 
(2) poorer effective linear resolution and therefore their VLBA cores apparently
contain a larger amount of linearly polarized jet emission.  The jet of 3C\,84
shows a complex circular polarization structure, similar to observations by Homan \&
Wardle five years earlier; however, much of the circular polarization seems to have 
moved, consistent with a proper motion of 0.06$c$.    
The jet of 3C\,273 also has several circularly polarized components, and 
we find their fractional circular polarization decreases with distance from the 
core.
\end{abstract}


\section{Introduction}
\label{s:intro}

Only a small fraction of the radio emission from
extra-galactic radio jets associated with Active Galactic Nuclei (AGN)
is circularly polarized; however, this weak component is a potentially 
important probe of the particle population 
and magnetic field structure of the jets.  Circular polarization (CP) 
can be produced either as an intrinsic component of the emitted
synchrotron radiation, or via Faraday conversion
of linear polarization into circular \citep{JOD77}.  If the intrinsic
mechanism dominates, jets must be a predominantly electron-proton
plasma and must contain a significant fraction of unidirectional
ordered magnetic field.  If the conversion mechanism dominates,
circular polarization probes the low energy end of the relativistic
particle distribution \citep[e.g.][]{WH03}, a key parameter in
studying the bulk kinetic luminosity of AGN jets and
their particle content \citep{CF93, WHOR98}.
Faraday conversion is expected to be the dominant mechanism on 
theoretical grounds \citep[e.g.][]{J88}; however, there is still 
only limited observational evidence supporting this view
\citep{WHOR98, HW04}.

Early integrated measurements of CP from AGN jets \citep{WdP83, K84}
revealed circular polarization to be only a very small fraction,
typically $< 0.1$\%, of their total flux.  However, the Very Long Baseline 
Array (VLBA)\footnote{The VLBA is operated by the National Radio Astronomy
Observatory, a facility of the National Science Foundation operated under
cooperative agreement by Associated Universities, Inc.}, has made it possible to 
study circular polarization from extragalactic jets at sub-milliarcsecond
resolution.  \citet{WHOR98} and \citet[hereafter HW99]{HW99} found circular 
polarization in cores of four powerful AGN jets at {\em local} fractional levels of
$0.3$ to $1$\% of Stokes-$I$ with the VLBA.  Circular polarization has since 
been detected with various instruments in other AGN jets \citep{RNS00, HAW01}
and a wide variety of other synchrotron emitting sources:  low-luminosity AGN such
as Sagittarius A* and M81* \citep*{BFB99,SM99,BBFM01,BFM02}, intra-day
variable sources \citep{MKRJ00}, and galactic micro-quasars 
\citep{Fend00,Fend02}.  

Despite the detection of the circular polarization in a variety of sources,
observations of circular polarization from AGN jets are fragmented and incomplete.  
A large temporal gap exists between
the integrated measurements of the 1970's to early 1980's and the high 
resolution VLBA measurements from the mid 1990s onward.  Only a few
AGN jets have high resolution, multi-epoch circular polarization 
observations that are matched in observing frequency and sensitivity. 
An even smaller number of sources have multi-frequency circular polarization
observations suitable for measuring a spectrum.  Additionally, the 
small, {\em ad hoc} samples studied to date have not allowed for an 
analysis of correlations between the appearance of strong circular 
polarization and other source properties.  While these previous studies
raised a number of interesting possibilities of how circular polarization
could be used to constrain the magnetic field properties and 
particle populations of AGN jets, a systematic survey of a large 
complete sample is necessary to make further progress.

We have exploited the VLBA's unique imaging capabilities by studying the
milliarcsecond-scale circular polarization of a complete AGN jet
sample at 15 GHz as part of the MOJAVE (Monitoring Of Jets in AGN with 
VLBA Experiments) program.  
The MOJAVE sample 
of 133 AGN is complete and flux limited on the
basis of VLBA-scale emission.  The selection criteria are described
in detail in \citet{LH05} which also reports first-epoch linear
polarization results; however, in brief, for inclusion in MOJAVE,
a source must have had a total 15 GHz VLBA flux density of at least
1.5 Jy ($> 2$ Jy for sources south of the celestial equator) at any
epoch from 1994--2003.  When completed, MOJAVE will have obtained
at least four epochs of full Stokes VLBA images at 15 GHz of all
sources in the sample.  In addition, multi-frequency follow-up 
programs will study the circular polarization spectrum of the strongly
circularly polarized sources identified by MOJAVE.

In this paper, we report only the first-epoch circular polarization
results from the MOJAVE program.  Future papers will explore other
aspects of this rich data set including multi-epoch circular polarization 
variability.  The details of our observations and calibration appear
in section 2.  We present and discuss our results in section 3, and 
our conclusions appear in \S{4}.

\section{Observations and Calibration}
\label{s:obs}
Each MOJAVE observing session targets 
18 sources over a 24 hour period, and this paper presents results drawn from 
the first eleven sessions, from 2002 May 31 through 2004 February 11.  
During each 24 hour session, observations of the 18 sources are broken up 
into highly interleaved scans, each approximately seven minutes in duration.  
The schedule is designed to maximize coverage in the $(u,v)$-plane; however, as 
discussed below, the highly interleaved nature of the observations also improves our
ability to calibrate the antenna gains to detect circular polarization.  We 
observed each source for approximately 65 minutes of total integration time. No additional 
calibrators were observed as all of our target sources were strong and 
compact.

The data were recorded at each VLBA antenna at a central observing frequency
of 15.366 GHz using 1-bit sampling and were correlated at the VLBA correlator 
in Socorro, NM.  The correlator output was 2 s integrations for all four 
cross-correlations (RR, RL, LR, LL).  The correlated data contained four 
intermediate frequencies (IFs) and 16 channels per IF, yielding a total bandpass 
of 32 MHz in both RR and LL.  Tapes of the correlated data were shipped to NRAO 
in Charlottesville, 
VA and Purdue University where they were loaded into NRAO's Astronomical Image 
Processing System (AIPS) for further processing.

The basic calibration procedure followed the standard methods outlined
in the AIPS cookbook and the procedure by HW99 for measuring
circular polarization with the VLBA using the {\em gain transfer} technique.  
Below we briefly describe our basic calibration procedure, in
\S{\ref{s:gains}} we discuss some improvements to the {\em gain transfer} 
procedure developed for these observation, and in \S{\ref{s:phase}}
we discuss some checks on our calibration procedure.

The data were loaded into AIPS using the VLBALOAD procedure.  The procedure
VLBACALA was then used to do a-priori amplitude calibration, running both the
ACCOR and APCAL tasks, including an atmospheric opacity correction.  The parallactic angles were
removed using the procedure VLBAPANG and the pulse calibration was applied
using VLBAPCOR.  The procedure VLBAFRNG was then used to perform fringe fitting.
All of our sources were strong enough to serve as their own fringe calibrator, and 
no special interpolation from strong to weak sources was necessary.  
We used the procedure CRSFRING to remove any residual multi-band delay
difference between the right and left-hand systems of the array.  Finally we ran
BPASS to apply a bandpass correction before averaging across the channels within
each IF and 'splitting' the data off into individual source files. 

Following a 2 s time-scale point-source phase calibration to increase 
coherence, the data were written out to the CALTECH VLBI program, DIFMAP 
\citep{S97}.  In DIFMAP, the data were averaged to 30 seconds and edited
in a station-based manner.  An edited version of the data was then saved without
any self-calibration applied.  We will refer to this version of the data as the 
edited/unself-calibrated data.  Normal self-calibration and imaging techniques 
were then used within DIFMAP to obtain the best Stokes-$I$ model for
each source.  

For each source, the best Stokes-$I$ model and the edited/unself-calibrated data 
were then read back into AIPS, and the data were self-calibrated against the model 
using the task CALIB in amplitude and phase (SOLMODE = 'A\&P') on a 30 second 
timescale.  The antenna feed-leakage term corrections from LPCAL (see below) were applied 
using 
SPLIT and the parameter DOPOL = 3 to guarantee that the corrections were applied 
to both the parallel and cross-hand correlations.  A final pass of CALIB self-calibrated 
the data again against the Stokes-$I$ model to remove any errors
by the initial self-calibration prior to D-term removal.  For circular polarization
calibration via {\em gain transfer} it is important that both passes of 
CALIB described above were performed only making the rigorous assumption\footnote{For 
the total intensity and linear polarization imaging described in \citet{LH05} we 
assumed there was no Stokes-$V$ in these CALIB steps by assuming
$RR = I_{model}$ and $LL = I_{model}$.} 
that $(RR+LL)/2 = I_{model}$.  This left as un-calibrated 
any gain offsets between the right (R) and left (L) hand feeds at each antenna.  
The {\em gain transfer} calibration described in \S{\ref{s:gains}} was 
used to remove these offsets.  After correction of the $R/L$ antenna gains, 
final images were made in all four Stokes parameters. 

It is important to note that the antenna feed leakage corrections were initially 
determined on each source individually using LPCAL, and the final corrections (which
were applied to the data) were determined by taking the median of these 
after removing obvious outliers.  Our typical feed leakage terms (D-terms) had amplitude of  
about 0.02 with the largest being 0.05, and we estimate that we have made these 
corrections accurate to about $\pm0.005$ .  Following HW99, we estimate any uncertainty
in our fractional circular polarization measurements, $m_c$, due to leakage 
correction errors to be of the order $\Delta m_c < 0.001 m_L + 1\times 10^{-5}$, where $m_L$ is the
fractional linear polarization of the map peak (typically a few percent or 
less).  Thus any uncertainty due to leakage feed corrections is very much smaller
than the uncertainty due to the gain calibration and can be neglected.

\subsection{R/L Gain Calibration} 
\label{s:gains}
As described in HW99 and \citet{HAW01}, the {\em gain transfer} technique
requires that we initially make no assumption about circular polarization during
self-calibration, i.e., we only assume $(RR+LL)/2 = I_{model}$ as discussed above.
This will leave as un-calibrated a small relative R/L complex gain ratio at each VLBA
antenna. The essential idea is to calibrate this gain ratio with a smoothed set of 
R/L complex antenna gains determined from all 18 sources in each 24 hour observing
session.

To determine the smoothed set of antenna gains that will ultimately be applied 
to the data, we first solve for the R/L complex gain ratios by 
self-calibrating each source assuming no circular polarization in that 
source ($RR = I_{model}$ and $LL = I_{model}$).   These solutions are not
applied to the data, but rather they are collected into a single calibration table 
which we smooth by using a six-hour running median with all sources equally 
weighted.  Solutions from sources with 
low gain SNR or very strong circular polarization (greater than 0.5\%) are 
prevented from contributing to the running median\footnote{Such strongly
polarized sources were very rare; however, when we did find one, we simply 
repeated the calibration with the source in question prevented from contributing
to the median gain solution}.  To account
for the possibility of smaller levels of real circular polarization showing up in the 
raw R/L antenna gains of individual sources, we subtract from the raw R/L gain ratios for 
each source a single mean offset (taken across all times, antennas, and IFs 
for that source) from the running median.  The sum of the applied offsets for all
sources in a given 24 hour observing session is constrained to be zero.  This constraint 
guarantees quick 
convergence of an iterative algorithm which generates a trial set of smoothed gains,
computes mean offsets for each source, takes those offsets out of the data, and then
repeats the process.  The constraint also underscores the fundamental assumption of
the gain transfer technique: that any single observing session will have an average 
circular polarization of zero across all the sources in the sample\footnote{AGN as a
whole have no preferred sign of circular polarization, and thus we expect the net
circular polarization from a large sample of AGN to be close to zero.}.  This assumption 
may introduce a small overall gain bias in any individual epoch; however, our Monte
Carlo method of estimating the uncertainty due to the gain calibration (described below) 
naturally accounts for this possibility.

We note that the procedure described above for accounting for sources with small levels
of real circular polarization differs from the procedure used in
HW99 and \citet{HAW01}.  In those papers, we prevented such sources
from contributing to the smoothed gains through an iterative process where the strongest
polarized sources were removed first, the smoothed gains recomputed, and the process 
repeated until no source contributing to the smooth gains appeared to have real circular 
polarization.  In those papers, we found the final results were insensitive to the
detailed steps of this process; however, the process was time consuming and difficult to
automate.  For the MOJAVE program, we needed a procedure that was automated, reproducible,
and could be easily tested by a Monte Carlo simulation we developed.  As a check that the
new procedure gave the same results as our previous methods, we reprocessed the antenna gain 
solutions from the 40 source observation analyzed in \citet{HAW01}.  For the circularly 
polarized sources in that experiment, we found an RMS R/L gain difference of $0.017$\% 
between our technique and the original.  This difference is approximately one-third of the 
estimated gain uncertainty of that experiment, verifying that the two techniques indeed 
give effectively the same results.

\subsubsection{Uncertainty in Measured CP due to Gain Transfer}
In the appendix of \citet{HAW01}, we laid out a prescription for estimating
the uncertainty in circular polarization measurements made by gain transfer.
Here we improve on that work by using a Monte Carlo simulation to test our
methods and obtain more accurate uncertainty estimates.  As in \citet{HAW01}, 
we focus on amplitude gain errors, as those 
dominate the uncertainty in the technique.  Phase errors will not contribute 
to first order for point-like sources, and for sources with significant extended 
structure, phase errors will produce anti-symmetric structure around the phase
center.  In section \S{\ref{s:phase}} we discuss possible phase 
effects and tests to constrain/resolve those issues.

The Monte Carlo simulation we developed generates 18 point sources (Most sources in 
our sample are dominated by a point source, e.g., \citet{LH05}.), and for each 
source generates a circular polarization 
with magnitude drawn at random from the distribution of observed
mean gain offsets in the real gain table.  The sign ($+$ or $-$) of the CP signal 
for each source is determined at random.  Finally, a Gaussian deviate representative of our
measurement uncertainty is added to each value.  Each of these generated CP values is 
then stored for comparison to the ``measured'' value at the end of a Monte Carlo trial.

To generate the gain table for the simulation, a random slowly-varying gain curve
is created for each antenna/IF with amplitudes and timescales similar to those 
in the real gain curves.  To this slowly-varying gain curve is added randomly-generated 
scan-to-scan variations of the same range of amplitudes as seen for the corresponding
source in the real gain table.  Finally, the generated CP values are added to the 
gain curves as offsets for each source.  This step is important to mimic the real gain 
table, which was generated from the assumption that each source has no circular 
polarization, so the real raw gain table contains 'corrections' to the circular 
polarization of each source in addition to the actual gain variations. 

Finally, this randomly-generated 'raw' gain table is processed by exactly the same 
procedure we used on the real gain curve.  A smoothed version of the raw table is 
produced by our standard procedures and compared with the raw gain table.  If the procedures
work correctly, the only differences between the raw and smoothed tables should be
the circular polarizations set at the beginning of the simulation trial.  Therefore, 
by comparing the two tables, we can estimate the uncertainty
due to the gain calibration procedure.  We repeat this simulation for 
1000 independent trials to generate an estimate of the uncertainty for each source.

To safe-guard against the possibility that an unusual combination of circularly 
polarized sources in our data has made our results more uncertain than the above
simulation indicates, we also ran a second version of the Monte Carlo described above.  
The only difference is that the initial circular polarization of each source in the 
simulation is forced to match the mean gain offset of the corresponding source in the real 
gain table (plus a Gaussian deviate to represent measurement uncertainty).  
To these values, we add a randomly-determined overall bias (the same for all 18 sources)
to account for the possibility that the circular polarizations in the sample do not 
average to zero.  The size of this gain bias is determined from the RMS of the mean 
offset values in the real gain table.  The larger of the uncertainties from the first and 
second version of the Monte Carlo simulations are used to represent the total gain uncertainty for a 
given source; however, in practice the two Monte Carlo simulations give very nearly the 
same estimate of uncertainty. 

\begin{figure}
\plotone{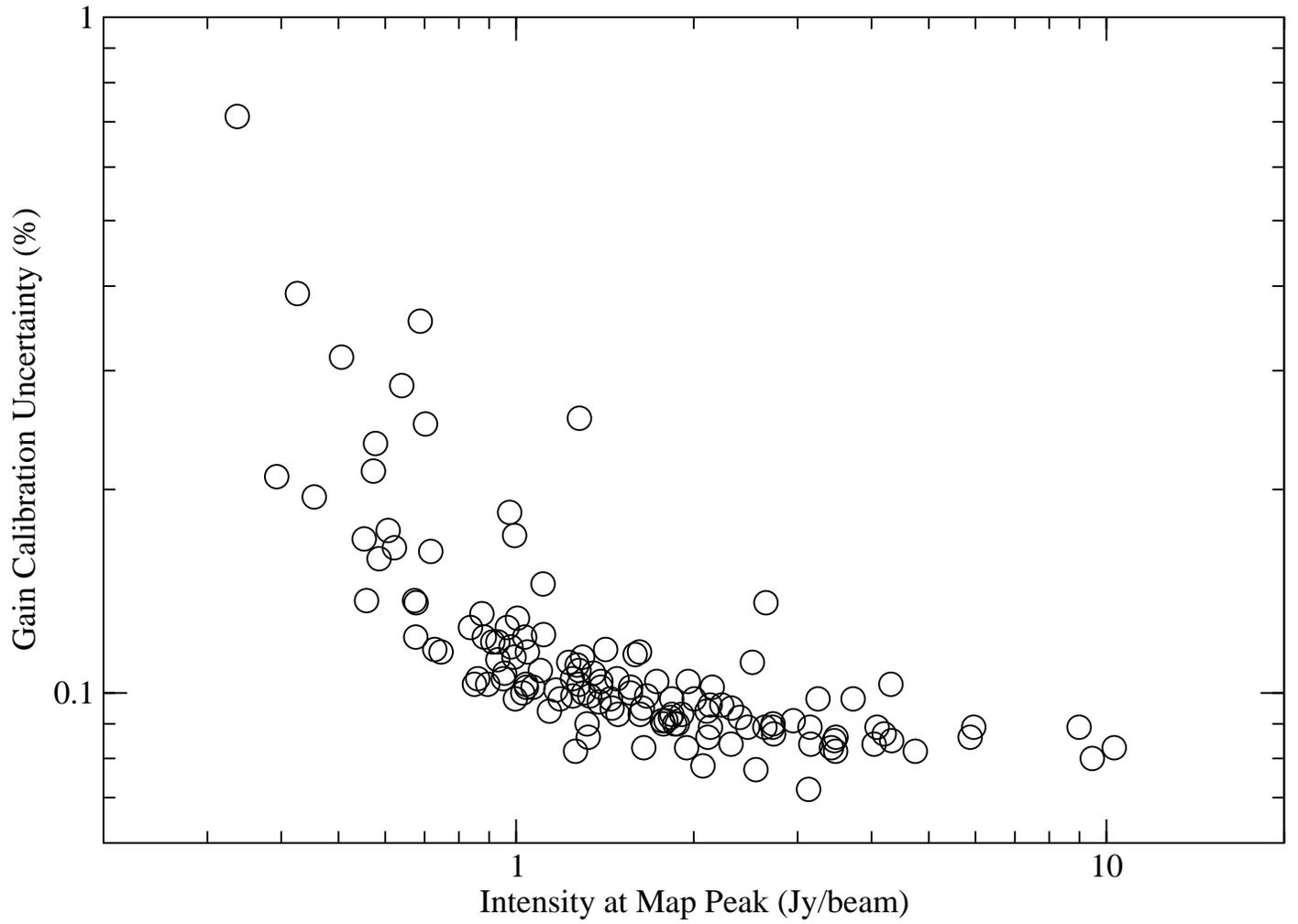}
\figcaption{\label{f:gains}
Gain calibration uncertainty for circular polarization (as a percentage of Stokes-$I$) as
a function of Stokes-$I$ intensity at the map peak for the 133 sources in our sample.
}
\end{figure}

Figure \ref{f:gains} shows a plot of uncertainty in fractional circular polarization 
measurements due to gain calibration as a function of the peak Stokes-$I$ flux for 
the 133 sources in the MOJAVE sample. The trend 
of larger uncertainties for weaker sources is expected and is driven by the larger 
self-calibration gain uncertainties in the weaker sources; however, it will be 
important to consider
this trend when interpreting the results in \S{\ref{s:res}}.  Note that most sources 
in the sample have uncertainty contributions from gain calibration very close to 
0.1\%, and given that the
gain uncertainties dominate the total uncertainty (the much smaller RMS map noise is
added in quadrature), our procedures are able to robustly detect CP signals as small
as 0.3\% of the Stokes-$I$ peak at the 3-$\sigma$ level. 

\subsection{Phase Calibration for Circular Polarization}
\label{s:phase}
Because there is no phase connection between sources in our sample (they are well separated
in the sky and fringe-fit separately), the gain calibration technique described above is 
only effective for calibrating the antenna amplitude gains.  We do allow the possibility of
a phase correction in the gain transfer technique (the median phase difference between
the right and left hands is treated in a similar fashion to the median amplitude ratio.); however
the applied corrections are generally $\lesssim 0.1$ degrees for each antenna/IF combination.

For effective relative R/L phase correction at each antenna we rely first on the pulse 
calibration corrections, then 
on fringe fitting and the initial phase calibration in CALIB to increase coherence before 
averaging the data in time in DIFMAP.  Each of these steps provides separate phase corrections 
for the left and right-hand feeds at each antenna.  The pulse calibration makes no assumptions 
about circular polarization in the source; however both fringe fitting and the initial phase 
calibration to increase coherence make their corrections assuming the source is point-like.  
Any real circular polarization on a point source will not appear as a gain error in the antenna 
phases (HW99), so for most sources in our sample phase corrections found in this fashion 
will remove only genuine R/L phase gain errors from the data.

\begin{figure}
\begin{center}
\includegraphics[width=5in]{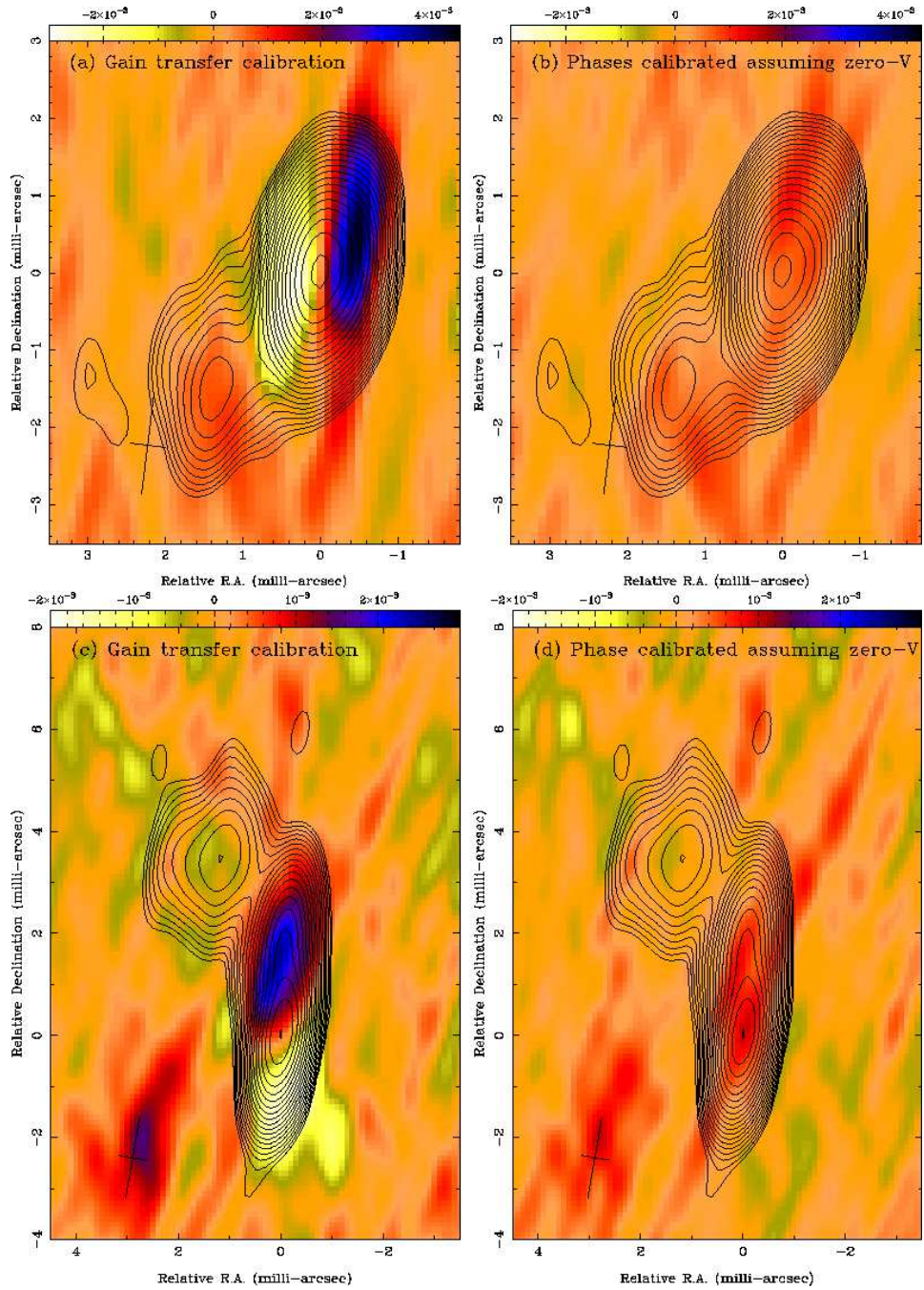}
\end{center}
\figcaption{\label{f:pz_nocp}
Circular polarization (color-scale given in mJy beam$^{-1}$) 
superimposed on Stokes-$I$ contours for 1213$-$172 (top panels)
and 2243$-$123 (bottom panels). The $I$ contours start at 1 and 2 mJy beam$^{-1}$ 
respectively for 1213$-$172 and 2243$-$123 and increase in steps of $\times\sqrt{2}$.
The FWHM dimensions of the
restoring beams are indicated by a cross in the lower left corner of each panel.
Left hand panels (a,c) show the results 
of straight gain transfer calibration which has a combination of real circular
polarization plus phase errors.  Right hand panels (b,d) show the additional 
effect of a round of phase self-calibration assuming no circular polarization 
in the data.  
For both sources, the additional phase calibration has corrected a phase 
shift in the data between the left and right hand data, and reveal that
the source has no real circular polarization.  
}
\end{figure}

If a source has significant extended structure in Stokes-$I$, the right and left hand systems of the array 
may receive slightly different solutions in the algorithm's attempt to find point source solutions 
to the right and left hand data separately.  Most likely this will generate a small difference in the
phase centers of the right and left hand data.  While this small difference will not affect
the Stokes-$I$ image, it may be enough to generate an anti-symmetric stripe of circular 
polarization across the strongest parts of the source. The effect is only apparent in the
circular polarization image because Stokes-$V$ is formed by taking the difference of the
right and left hand data ($V = (RR-LL)/2$), whereas Stokes-$I$ is formed from their average.  
See panels (a) and (c) of Figure \ref{f:pz_nocp} for examples of this effect.  

To account for this effect and other possible phase errors in our data, we have generated two 
circular polarization images of each source.  The first image is generated from
the gain transfer calibration alone and may contain unaccounted-for phase errors.  The second
image is generated after an additional round of self-calibration of the phases alone.  During this 
self-calibration of the phases, we assume there is no circular polarization in the data so
both the $RR$ and $LL$ data are compared to $I_{model}$ to find the antenna phase corrections
for the right and left-hand feeds separately.  This self-calibration will very effectively 
remove any genuine R/L phase gain errors remaining in the data, as shown in panels (b) and (d) of
Figure \ref{f:pz_nocp}.  The large majority
of the sources in our sample are core-dominated and thus show no significant differences 
between the images produced before and after this phase self-calibration; however, an 
important minority of sources with significant extended structure do show differences 
(usually both in their noise levels and in their apparent circular polarization) between 
the two sets of images.

\begin{figure}
\begin{center}
\includegraphics[width=5in]{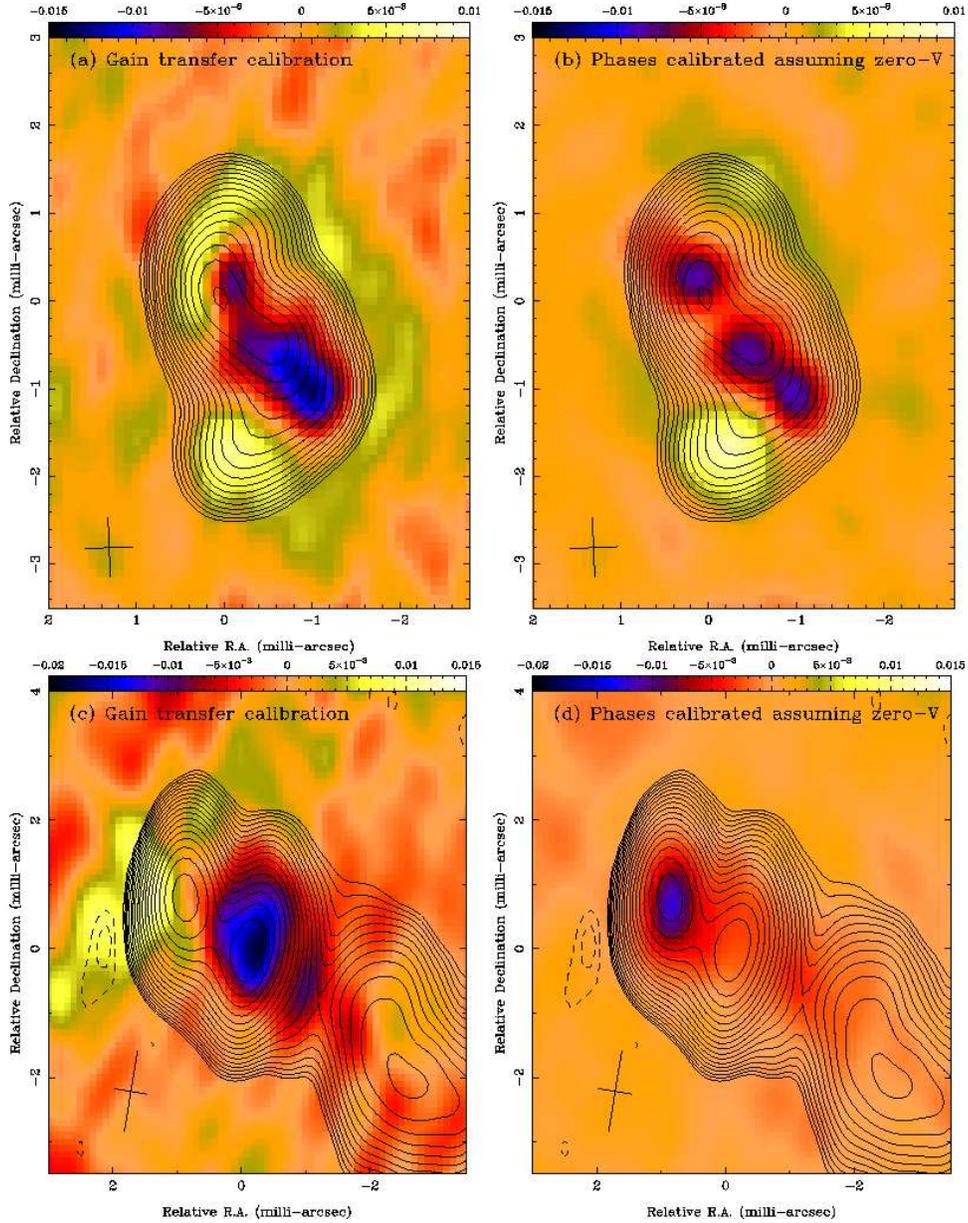}
\end{center}
\figcaption{\label{f:pz_cp}
Circular polarization (color-scale given in mJy beam$^{-1}$) 
superimposed on Stokes-$I$ contours for 3C\,84 (top panels)
and 3C\,273 (bottom panels).  The $I$ contours for both sources
start at 10 mJy beam$^{-1}$ and increase in steps of $\times\sqrt{2}$. 
The FWHM dimensions of the
restoring beams are indicated by a cross in the lower left corner of each panel.
Left hand panels (a,c) show the results 
of straight gain transfer calibration which has a combination of real circular
polarization plus phase errors.  Right hand panels (b,d) show the additional 
effect of a round of phase self-calibration assuming no circular polarization 
in the data.  For 3C\,84, panel (b), the additional phase calibration has corrected the phase 
errors and revealed the circular polarization of the source in greater
detail, although the main features appear quite similar to panel (a).
For 3C\,273, panel (d), the additional phase calibration has corrected the phase
error and reveals that the real circular polarization in the source is 
on the jet core and not on the jet component.  Note that the images shown
here for 3C\,273 are from our second MOJAVE epoch on this source (August 28, 2003), 
but they are included here as an illustrative example. 
}
\end{figure}

Figure \ref{f:pz_cp} illustrates two cases where the antenna phase errors
have been removed to reveal real circular polarization.  It is important to know to what
extent (if any) the resulting circular polarization in these images has been modified by
the phase calibration assuming zero circular polarization.  In general, phase differences
between RR and LL due to real circular polarization do not show up in a consistent antenna-based
fashion (as would phase gain errors); however it is possible that some combination of 
antenna-based phase gains could be found by the self-calibration algorithm which minimizes 
these real phase differences and thus modifies the real circular polarization of the source. 

We have tested for this effect by (1) removing all circular polarization
from a real source (by setting $RR=LL=I$), (2) adding known amounts of circular 
polarization, and then (3) self-calibrating the phases of that data assuming no 
circular polarization.  We have done this for several sources from our observations, 
and in general we find that the known circular polarization is not modified very 
much by this procedure.  
We usually see some reduction in amplitude from a few percent up to about 10\% of the 
known CP value, and the location of the CP peak may be shifted a fraction of a beam-width to 
better align with the map peak (assuming it is near the map
peak to begin with).  In some sources with a broad core-region (core $+$ barely resolved 
component), a single component of CP is spread out by this self-calibration to encompass 
the whole region.  
Large shifts in CP location (more
than a half a beam-width) or large reductions in amplitude (more than 20\%) do not seem 
possible with this procedure.  For example, our tests on the images in Figure \ref{f:pz_cp}
show that real circular polarization that began on the first jet component would remain there
after phase self-calibration and would not be shifted to the core; thus, we can be confident
that the real circular polarization in this epoch of 3C\,273 is actually associated with
the core as revealed in panel (b).
   
\section{Results and Discussion}
\label{s:res}

Our first-epoch circular polarization measurements are summarized in Tables \ref{t:cp_tab1}, 
\ref{t:cp_tab2}, and \ref{t:cp_tab3}.  Table 1 lists the linear and circular polarization 
measurements at the peak of each map.  In most cases the Stokes-I map peak corresponds 
to the unresolved VLBI core; however, in a handful of cases the core and map peak do not 
coincide.  In most of these cases, the peak appears at the location of a strong component 
within the jet, and we take the core to be the compact component at the base of the jet.
The core of 0923$+$392 is taken at the location identified by \citet{AMZ93}.  The core of 0316$+$413 
(3C\,84) is a special case and is discussed further in \S{\ref{s:sources}}.
The off-peak jet cores are listed in Table 2, along with measurements of their linear
and circular polarization at the core location.  Table 3 lists the circular
polarization measured on all jet components in our sample with peak fluxes stronger than 
300 mJy beam$^{-1}$.  We chose 300 mJy beam$^{-1}$ as a reasonable cutoff for the jet component 
analysis as our typical uncertainty levels would still allow very strong circular 
polarization of $m_c = 0.5$\% to be detected on such a component.

The linear polarization was always measured at the location of the Stokes-I peak (or at 
the location of the VLBI core or jet component in the cases of Tables 2 and 3 respectively).
Limits are given for the linear polarization when the measured value does not exceed five 
times the map RMS noise.  Circular polarization was also measured at the location of the
Stokes-I peak (or at the location of the VLBI core or jet component for Tables 2 or 3) 
unless there was a CP peak within two pixels of the Stokes-I measurement location.  In that
case, the value of the CP peak was used instead.  This procedure was used because with our 
phase calibration (see section 2), there may be a small positional shift in the 
location of the circular polarization relative to the Stokes-I.  The raw circular polarization
measurements and their uncertainties are given in Tables 1, 2, and 3.  The uncertainties
includes both the gain transfer uncertainty from our Monte Carlo simulation (see \S{2})
and the RMS noise\footnote{The RMS noise was calculated over a large number of pixels away
for source structure.  For each pixel, we used largest value of CP within 2 pixels of
that location.  This procedure increases our estimate of the RMS noise, but it is consistent
with the procedure we use for measuring CP from a Stokes-I peak.} added in quadrature.  It is important
to note that the fractional circular polarization values reported in the Tables include 
both measurements and limits.  While we do actually measure a CP value in every case, most 
of these measurements are not significant, and thus we have chosen a $2\sigma$ cutoff, below 
which we report a limit on the value of $|m_c|$.  The quoted limits on $|m_c|$ are the larger of 
$\left(|\mathrm{V}| + 1\sigma_\mathrm{V}\right)$ or $2\sigma_\mathrm{V}$.  The raw values of $m_c$ 
in the cases that are less significant than $2 \sigma_\mathrm{V}$ can be computed from 
the $V$ and $I$ intensities reported in the Tables.

We note that we have applied exactly the same procedures to the cores and jet components in 
Tables 2 and 3 as we applied to the map peak values; and, as argued in \S{\ref{s:gains}}, the 
interpretation of the gain uncertainties is most robust for the map peak.  Therefore, we are
somewhat less confident of the uncertainties reported in Tables 2 and 3; however, we believe
that the gain uncertainties will still accurately account for the possibility of any overall 
gain bias in the CP map. The gain uncertainties do not account for the possibility of
random (but not net) gain errors creating small amounts of spurious circular polarization off
the map peak; however, the RMS noise level in the map naturally takes this into account.  We notice 
no particular trend for map noise to be scattered preferentially onto weaker source structure, 
and we therefore believe that the combination of gain uncertainties and map RMS noise provides 
a reasonable uncertainty estimate for the off-peak circular polarization reported in Tables 2 
and 3.     

\subsection{Circular Polarization on VLBI Cores}
\label{s:cores}

\begin{figure}
\begin{center}
\includegraphics[width=2.5in]{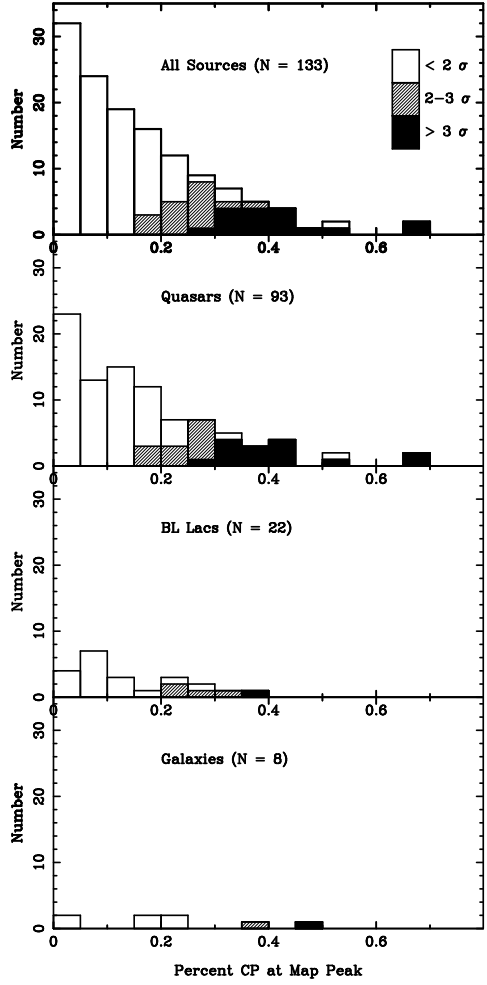}
\end{center}
\figcaption{\label{f:core_dist}
Histograms of the absolute value of the fractional circular polarization at the 
location of the VLBI jet core with each of the bottom three panels
corresponding to a different optical classification.  The significance 
of each measurement is indicated by open, hash, or solid fill styles 
for $< 2\sigma$, $2$--$3\sigma$, and $>3\sigma$ values respectively.  Note that no limits
appear on these histograms.  For sources with limits in Table 1 or 2, we 
have substituted the fractional circular polarization at the core location
calculated from the $V$ and $I$ values reported in the table.
}
\end{figure}

Figure \ref{f:core_dist} shows the distribution of core circular polarization for all the sources
in our sample.  For the purposes of this figure we use measured values rather than limits
in every case.  We detect circular polarization in 17 out of 133 jet cores at the $3\sigma$ level
or higher significance.  Of the 115 sources with overall uncertainties less than $0.15$\%, we 
measure circular polarization stronger than $0.3$\% of the local Stokes-I in 17 of those 
sources (all but two of those at $\geq 3\sigma$ significance).  The existence of strong
circular polarization ($\geq 0.3$\%) in about 15\% of our sample is somewhat higher than
the rate of strong circular polarization observed by \citet{HAW01} with the VLBA at 5 GHz.  
They found CP stronger than 0.3\% in only 2 out of 36 sources (6\%); however, these rates of 
strong circular polarization detection are compatible according to Poisson statistics.

We examined the circular polarization distribution between source types \citep[using the
optical classifications given by][]{LH05}, and a Gehan's
Generalized Wilcoxon  test from the ASURV package \citep{LIF92} shows no significant difference 
between the distributions for quasars and BL Lac objects. 
Unfortunately there are simply too few galaxies in the MOJAVE sample for a robust test 
of their distribution.  A Gehan's test also shows no significance difference between 
the EGRET and non-EGRET detected sources \citep[as defined in][]{LH05} in terms of
their CP levels.

Due to the large number of limits in our sample, we used a Kendall-Tau test for censored data 
\citep{LIF92} to test for correlations between circular polarization and various other source properties.
Table \ref{t:tau} collects the results of these tests.  The Kendall-Tau test gives the probability that
{\em no correlation} exists between fractional circular polarization, $|m_c|$, and the property
in question.  Small values in the Table indicate that a correlation may exist, and we take $p < 0.05$
as suggestive of a correlation.  Of the twenty properties tested in Table \ref{t:tau}, we find 
only three with $p \lesssim 0.05$: 15 GHz jet flux ($S_{jet}$), 15 GHz core/jet flux ratio ($R_{jet}$), 
and 15 GHz core/total flux ratio ($R_{total}$) as defined by \citet{LH05}.  These three properties 
are closely related through the prominence of the VLBI scale jet, so we really only have one property 
with a possible correlation at the $p < 0.05$ level, consistent with pure chance.  Removing 3C\,273 from 
consideration raises $p$ to $0.12$ for $S_{jet}$ and $0.09$ for $R_{total}$, so we don't find
strong evidence that circular polarization is correlated with any of these properties.

While Galactic latitude, $|b|$, is not correlated with circular polarization, it is interesting
that of the 12 sources near the galactic plane ($|b| < 10$ degrees), none of them have
circular polarization at $\geq 2\sigma$.  Treating these sources and those with $|b| > 10$ degrees as 
separate samples, a Kolomogorov-Smirnov gives only a 5.3\% chance they are drawn
from the same distribution.  Again, this 
low probability may be spurious as there seems little reason the galactic plane should 
interfere with circular polarization signals.  In particular, any galactic Faraday 
rotation should have no noticeable effect on circular polarization \citep[e.g.,][]{HAW01}.  

One property that might naturally be expected to correlate with core circular polarization is
core linear polarization.  In the most common models of circular polarization production \citep{WH03},
both linear and circular polarization increase with increasing field order; however, the two may not
be correlated in a few scenarios: (1) if circular polarization is produced through a high internal 
Faraday depth model \citep[e.g.,][]{RB02,  BF02}, (2) if the linear polarization is externally 
depolarized \citep[e.g.,][]{HAW01}, or (3) if the circular and linear polarization we observe are 
produced in two distinct regions of the jet but are unresolved in what we see as the VLBI core. 

\begin{figure}
\begin{center}
\includegraphics[width=2.5in,angle=-90]{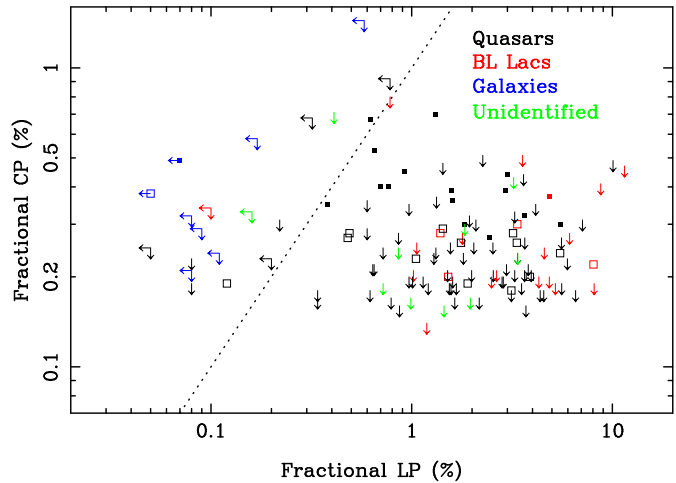}
\end{center}
\figcaption{\label{f:core_mcvml}
Plot of the absolute value of the fractional circular polarization versus fractional linear 
polarization for jet cores.  Black points represent quasars, red points
are BL Lac objects, blue points are galaxies, and unidentified
objects appear in green.  Solid squares are of $\geq 3\sigma$ significance,
open squares are of $2$--$3\sigma$ significance, and arrows represent
upper limits.  The dashed line indicates $|m_c| = m_l$. 
}
\end{figure}

Figure \ref{f:core_mcvml} is a plot of absolute fractional core circular polarization, $|m_c|$, versus 
fractional core linear polarization, $m_l$.  There is no clear relation between the two, and the 
obvious lack of 
correlation between linear and circular polarization is similar to the 5 GHz VLBA results of \citet{HAW01}, 
and the integrated 5 GHz measurements from the ATCA by \citet{RNS00}.  The persistence of no trend
at higher frequency suggests that the complicating factors (lack of 
resolution/Faraday effects) are to some degree self-similar and scale-up with observing frequency.  

Although there is no correlation between linear and circular polarization for jet cores, Figure
\ref{f:core_mcvml} shows that for quasars and BL Lac objects, linear polarization almost always 
exceeds circular polarization.  This does not seem to be the case for radio galaxies, which 
generally have weak linear polarization.  \citet{BBFM01} have noted
that the nearby very low luminosity radio sources Sgr A* and M81* are circularly polarized but
show no linear polarization. These authors have suggested that the $m_c/m_l$ ratio is distinctly
different for low luminosity AGN ($m_c/m_l >> 1$) and high luminosity AGN ($m_c/m_l << 1$).  
This relation was
tested further by \citet{BFM02}, who failed to find additional LLAGN with circular polarization;
however, if we consider the lowest redshift (and hence lowest luminosity) sources in our sample 
(which are admittedly quite different in character from the Sgr A* and M81*), we find that
$m_c/m_l >> 1$ for M87 (1228$+$126) and 3C\,84 (0316$+$413), the only two radio galaxies 
for which we can do the comparison.
Both M87 and 3C\,84 are known to have strong external Faraday screens which can depolarize the 
linear polarization \citep{W71, ZT02}, and in fact, all the radio galaxies in the MOJAVE sample 
have very low levels of core
linear polarization, likely due to external depolarization.  Higher redshift sources may
have smaller $m_c/m_l$ ratios not because a different production mechanism for CP, but rather because 
either (1) they have less depolarization, and/or (2) we study them with less effective linear 
resolution and therefore polarization from more of the jet contributes to what we see as the "core".  
If (2) is the case, it also naturally explains the lack of correlation between $m_c$ and $m_l$ in 
Figure \ref{f:core_mcvml}.  

\subsection{Circular Polarization of Jet Components}

Table 3 lists circular polarization measurements for all strong jet components in our sample with
peak Stokes-I $\geq 300$ mJy beam$^{-1}$.  Of the 33 jet components that meet this criteria, we detect
circular polarization in six (18\%) of them at the $3\sigma$ level or higher significance.  
Five of these six
detections are on the relatively nearby sources 3C\,84 (0316$+$413) and 3C\,273 (1226$+$023) which 
are discussed in more detail
in section \S{\ref{s:sources}}, and we note that 3C\,84 is a special case as the 
distinction between the VLBI core and jet is not well defined and the circular polarization measured is 
unusually strong ($\gtrsim 1$\%).  For these reasons, we must be cautious about drawing conclusions 
about the circular polarization characteristics of jet components in general; however, it is clear
that strong jet components can be circularly polarized, and the levels of circular polarization found in 
strong jet components are not clearly different from jet cores.  

\subsection{Discussion of Selected Sources}
\label{s:sources}

Here we discuss a few sources in detail, and we also compare our first
epoch results with previous VLBA observations of circular polarization 
(HW99; \citet{HAW01}; \citet{HW03}; 
\citet{HW04}) and from historical integrated measurements (1970s through 
early 1980s: \citet{WdP83}, hereafter WdP83; \citet{K84}, hereafter K84; 
\citet{AAP03}).  Because the latter are at lower frequency
(typically 5 GHz and below) and are integrated measurements, we only 
compare to the historical circular polarization results in those cases 
where the source was repeatedly (three times or more) detected in 
circular polarization.  

Because there is a large overlap of our sample with the sources observed 
at 5 GHz with the VLBA on 1996 December 15 by \citet{HAW01}, we list results 
from the latter study alongside our 15 GHz results in Table \ref{t:ba18}. We 
discuss some of the sources common to both of these studies in more detail 
below.  The large number of limits from both programs makes the results in 
Table \ref{t:ba18} difficult to interpret; however, it appears that CP of the 
core can vary with time and/or frequency.

\paragraph{0316$+$413 (3C\,84):}

\begin{figure}
\begin{center}
\includegraphics[width=3.5in]{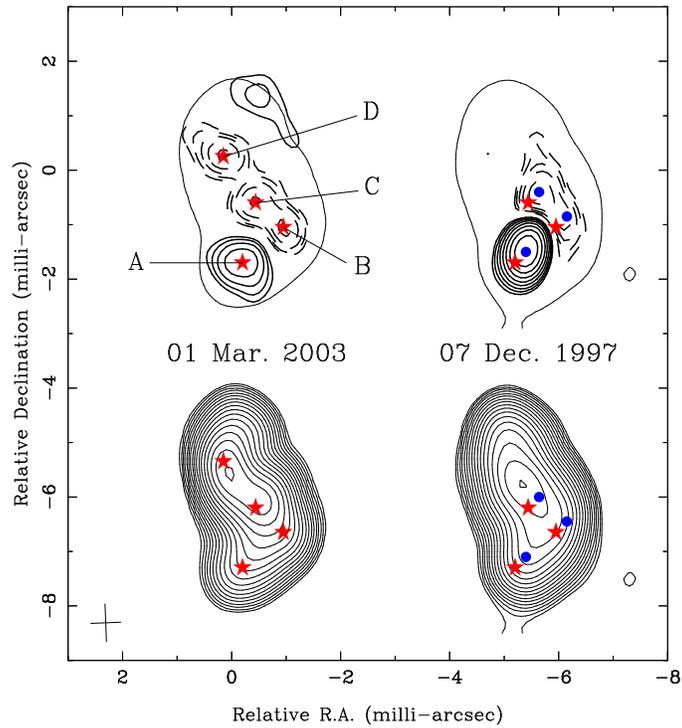}
\end{center}
\figcaption{\label{f:3c84_comp}
Circular polarization from the central 5 milli-arcseconds of 
3C\,84.  The left-hand panels show our March 2003 observations, 
and the right-hand panels show the observations by \citet{HW04}
from December of 1997.  The images from both epochs have been
convolved with an identical beam of 0.69$\times$0.55 mas$\times$mas,
plotted as a cross in the lower left hand corner of the plot.
The top images are circular polarization
contours beginning at $\pm 2$ mJy beam$^{-1}$ and increasing in steps of
$\times\sqrt{2}$.  Dashed contours represent negative circular 
polarization.  A single Stokes-$I$ contour shows the registration
of the CP within this region.  The bottom images are the Stokes-$I$
contours (beginning at $\pm 10$ mJy beam$^{-1}$ and increasing
in steps of $\times\sqrt{2}$) from the same region.  The four circularly
polarized components from March 2003 are labelled with letters and 
their positions are marked with red stars.  The three blue dots 
on the right hand image have the same relative positions as 
the red stars corresponding to components A--C, only they have
been shifted by 0.2 milli-arcseconds to the west and by 
0.2 milli-arcseconds to the north.
}
\end{figure}

The left-hand panels of Figure \ref{f:3c84_comp} show the central 5 milli-arcseconds
from our March 2003 MOJAVE observations of 3C\,84.  The image reveals four distinct 
spots of circular polarization within this central region of the source. 
Without a clear component to identify as the core, we have taken spot "D" as 
the location of the VLBI core. \citet{HW04} found that even the locations
corresponding to spots "C" and "B" were optically thick in December of 1997, so 
the precise definition of the jet core in this source is not clear.  With only
one frequency, we don't know where the $\tau\sim 1$ surface is in 2003, so it 
is possible that all three spots of negative circular polarization: B, C, and 
D, may be part of an extended optically thick region; however, for the purposes
of this paper, we have chosen to treat spots B and C as locations in the
jet, along with spot A.

Both spots D and C are similarly circularly polarized at the $-0.4$\% level; however,
the strength of the circular polarization increases sharply in spot B to  
$-1.4$\%.  A little further out in the jet, but at quite a different position
angle, spot A is more than $2$\% circularly polarized with {\em positive} 
circular polarization.  \citet{HW04} attribute this change in sign to
the stark change in opacity they observe between spots B and A.

The circular polarization we observe is very similar to that seen
by \citet{HW04} in observations more than 5 years earlier; however,
spot D is entirely new and was not present in December of 1997.  A version of
the 15 GHz image from \citet{HW04} (convolved with a matching beam to
our MOJAVE observations) appears in the right-hand panels of Figure
\ref{f:3c84_comp}.  The 1997 image is registered with the 2003 image by 
matching contours on the northern-most end of the central region shown here. 
The red dots in the right-hand panels of Figure \ref{f:3c84_comp} have
the same relative positions as the blue stars which mark the locations of
spots A, B, and C in 2003, but they have been shifted by 0.2 milli-arcseconds
to the North and by 0.2 milli-arcseconds to the West.  It is clear that
the {\em relative} positions of these circularly polarized spots has not
changed significantly in 5$+$ years; however, they do appear to move outward
together at a sub-luminal rate of roughly $0.06c$, almost directly 
South-East.  This speed is, of course, dependent on our alignment of
the images; however, it is similar to the speed found by \citet{DKR98} at 43 GHz 
for two of the three component they studied within this region, although 
those two components were moving almost due South.  

\citet{HW04} found spot A to have $m_c = +3.2\pm0.1$\% which is somewhat
larger than the $m_c = +2.3\pm0.3$\% we measure five years later.  The intensity
at the location of spot A has faded by a factor of three, from 1.2 Jy beam$^{-1}$ to 
0.4 Jy beam$^{-1}$, so presumably the region has become more optically thin.  Multi-frequency
observations are necessary to test if this is indeed the case, but if so, it is 
remarkable that this location has remained as strongly circularly polarized as 
it has.  \citet{HW04} argued that Faraday conversion is responsible for the
high levels of circular polarization detected here.  Faraday conversion 
is strongly dependent on optical depth, and we would eventually expect a
significant decline in circular polarization from this mechanism if this 
region of the jet is becoming increasingly optically thin \citep[e.g.,][]{WH03}. 
It will be particularly interesting to follow the changes in spot A in future
MOJAVE epochs.

\citet{HW04} found spots B and C to have $m_c = -0.7\pm0.1$\% and 
$m_c = -0.6\pm0.1$\% respectively at 15 GHz.  Spot B has essentially doubled
in circular polarization percentage in 5 years, and Spot C has not 
changed significantly within our uncertainties.  Neither location has 
changed much in Stokes-$I$ intensity.

If the alignment between the epochs is correct, the lack of large changes in 
the circular polarization distribution between spots A, B, and C is surprising 
given their motion.  The magnetic field order and particles responsible for 
producing this circular polarization must be in motion, however, this motion 
seems to have preserved the particular field characteristics responsible for 
the strong levels of circular polarization we observe.

\paragraph{0430$+$052 (3C\,120):}
We find a limit of $|m_c| < 0.29$\% in the core at 15 GHz in February of 2003.
HW99 also found limits between 0.2 and 0.3\%  in the core at 15 GHz throughout
1996.  Both signs of circular polarization (with a clear preference
for positive CP) were detected intermittently in the historical integrated 
measurements (WdP83 and K84) at 8.9 GHz and below.

\paragraph{0528$+$134:}
We find a limit of $|m_c| < 0.18$\% in the core at 15 GHz in February of 2003.
HW99 detected multiple epochs of significant circular polarization at about the
$m_c = -0.5$\% level in the core at 15 GHz during 1996.  During 1996, 
0528$+$134 was undergoing a large outburst, peaking with an integrated 
core flux 4.6 times larger than observed in February 2003. 

\paragraph{0607$-$157:}
We find a limit of $|m_c| < 0.21$\% in the core at 15 GHz in March of
2003. \citet{HW03} measured only $m_c = -0.23\pm0.12$\% in the core at 15 GHz in
January of 1998, but they simultaneously detected $m_c = -0.65\pm0.07$\% in
the core at 8 GHz.  \citet{HAW01} detected circular polarization in core of 
$m_c = -0.18\pm0.05$ \% at 5 GHz thirteen months earlier in 
December of 1996.  \citet{AAP03} found strong circular polarization,
$m_c = -0.23\pm0.04$\% at 4.8 GHz in integrated measurements with the
UMRAO during the period 2001--2002.

\paragraph{0735$+$178:}
We measure $m_c = -0.30\pm0.11$ \% in the core at 15 GHz in November of 
2002.  HW99 did not detect core circular polarization at 15 GHz at any epoch 
during 1996 with a typical limit of $|m_c| < 0.4$\%.

\paragraph{0851$+$202 (OJ287):}
We measure $m_c = -0.20\pm0.08$ \% in the core at 15 GHz in October of 2002.
HW99 did not detect core circular polarization at 15 GHz at any epoch during
1996, with a typical limit of $|m_c| < 0.2$\%.  \citet{HAW01} also found a limit of 
$|m_c| < 0.10$\% for the core circular polarization at 5 GHz in December of
1996.  In older integrated measurements during the period 1978--1983, 
\citet{AAP03} found strong CP at 4.8 and 8.0 GHz with $m_c = -0.37\pm0.12$\%
and $m_c = -0.22\pm0.03$\% respectively. 

\paragraph{0923$+$392 (4C 39.25):}
We find a limit of $|m_c| < 0.18$\% in the core at 15 GHz in October of 2002.  
In older integrated measurements during the period 1978--1983, 
\citet{AAP03} found strong CP at 4.8 GHz with $m_c = -0.21\pm0.05$\%.

\paragraph{1127$-$145:}
We measure $m_c = -0.24\pm0.10$ \% in the core at 15 GHz in October of 
2002.  Negative circular polarization was repeatedly detected at 5 GHz and
below in the historical integrated measurements cataloged by WdP83 and
K84.

\paragraph{1222$+$216:}
We find a limit of $|m_c| < 0.35$\% in the core at 15 GHz in February of 2003.
HW99 also found limits between 0.1 and 0.2\%  in the core at 15 GHz throughout
1996.

\begin{figure}
\begin{center}
\includegraphics[width=2.5in]{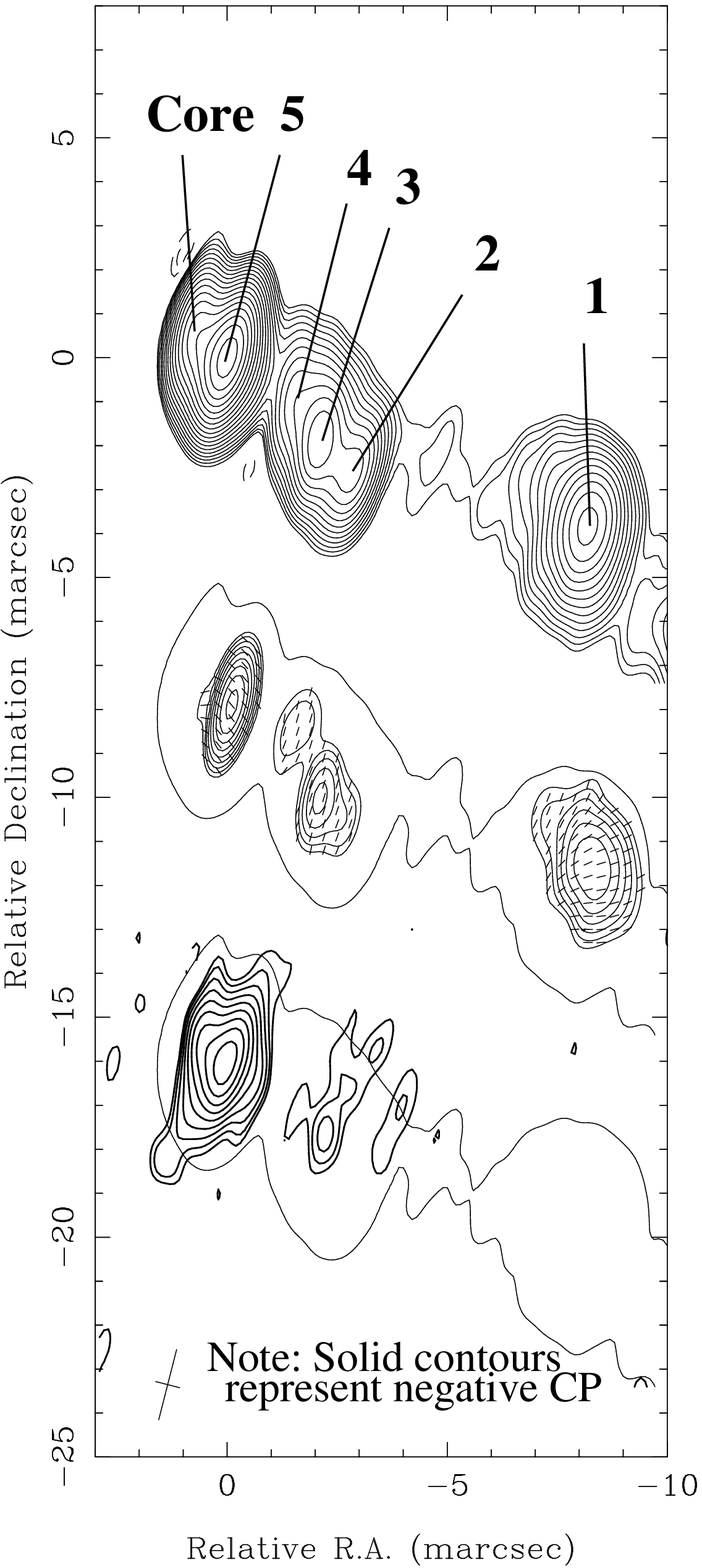}
\end{center}
\figcaption{\label{f:3c273cp}
Stokes-$I$ contours (top), linear polarization (middle), and circular polarization (bottom)
images of the inner 10 milli-arcseconds of the jet of 3C\,273 in October of 2002.  
A single Stokes-$I$ contour appears around the linear and circular polarization images
to show registration.  The Stokes-$I$ and linear polarization contours begin at $20$ mJy beam$^{-1}$,
and the circular polarization contours begin at $\pm2$ mJy beam$^{-1}$.  All contours increase in
steps of $\times\sqrt{2}$.  
In this image, negative circular polarization is indicated by
solid contours.  
The FWHM dimensions of the common
restoring beam is indicated by a cross in the lower left corner of the image.
Electric vector position angles (EVPA) for the linear polarization are 
indicated by tick marks in the middle image.  The locations of circular polarization 
measurements are indicated on the Stokes-$I$ map.
}
\end{figure}

\begin{figure}
\begin{center}
\includegraphics[width=3.5in]{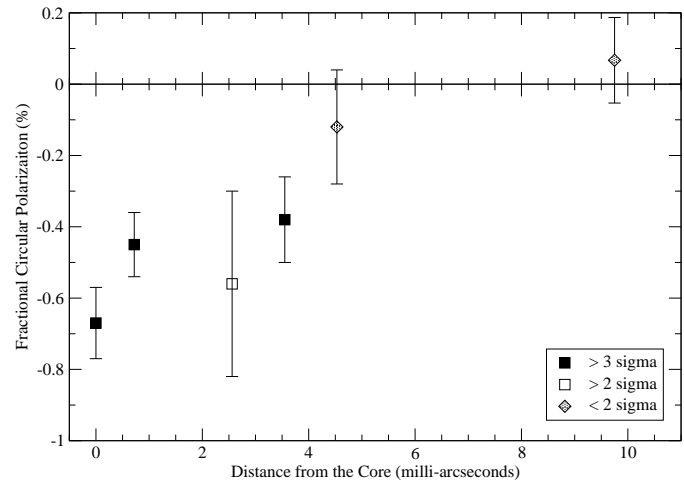}
\end{center}
\figcaption{\label{f:3c273_cpvr}
Plot of fractional circular polarization as a function of distance from 
the jet core for 3C\,273.  Note that the circular polarizaiton in 3C\,273 predominantly
has a {\em negative} sign and becomes weaker (going to smaller absolute values)
at greater distances from the core.  For the purposes of this plot,
we have plotted all the values with their corresponding error bars, including
those of $\lesssim 2\sigma$ significance. 
}
\end{figure}

\paragraph{1226$+$023 (3C\,273):}
Figure \ref{f:3c273cp} shows the first 10 milli-arcseconds of the jet of 3C\,273
at 15 GHz in October of 2002.  The top panel is Stokes I, the middle panel is
linear polarization, and the bottom panel is circular polarization.  
Positions where circular polarization measurements or limits were taken
are marked and numbered on the Stokes I map.  Figure \ref{f:3c273_cpvr} shows
the fractional circular polarization at these locations as a function of 
distance from the jet core.  The absolute fractional level of the circular 
polarization  appears to decrease as distance from the core increases.  This 
decrease could be tied to decreasing opacity along the jet or a decrease in poloidal
magnetic field strength.  

The core of 3C\,273 was found by HW99 to have $m_c \simeq -0.5$\% at three epochs during
1996.  The circular polarization appeared with the emergence of a new jet
component.  The core was very opaque at this time, and HW99 showed that the 
appearance of the CP was tied to the flattening of the core's spectral 
index due to the emergence of the new component.  As cataloged in WdP83
and K84, 3C\,273 was repeatedly detected in integrated measurements 
during the 1970s and early 80s at 8 GHz and below where it had a consistently
negative sign of circular polarization.

\begin{figure}
\begin{center}
\includegraphics[width=3.5in, angle=-90]{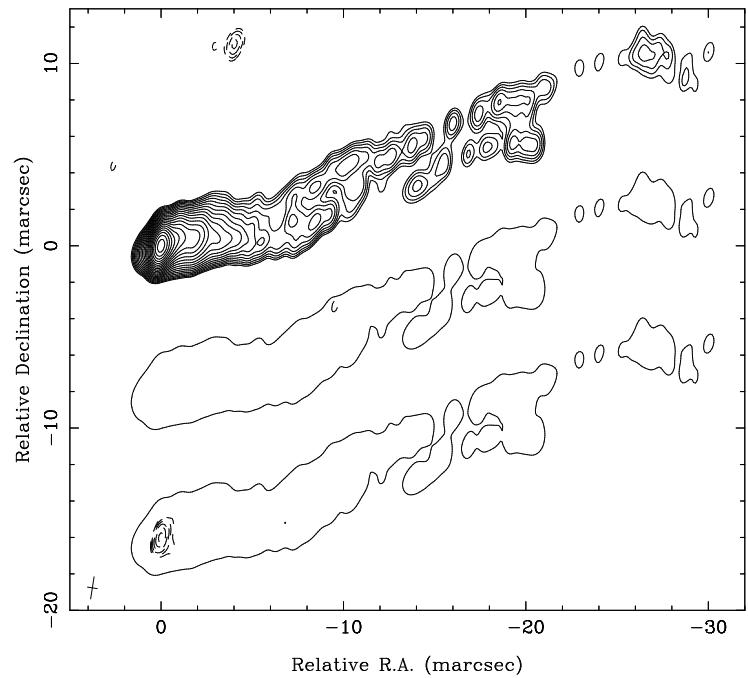}
\end{center}
\figcaption{\label{f:m87ipv.ps}
Stokes-$I$ contours (top), linear polarization (middle), and circular polarization (bottom)
images of the VLBI jet of M87 in February of 2003.  A single Stokes-$I$ contour appears 
around the linear and circular polarization images
to show registration.  All contours begin at $\pm 1$ mJy beam$^{-1}$ and increase in
steps of $\times\sqrt{2}$. Negative circular polarization is indicated by dashed contours.
The FWHM dimensions of the common
restoring beam is indicated by a cross in the lower left corner of the image.
}
\end{figure}

\paragraph{1228$+$126 (M87):}
Figure \ref{f:m87ipv.ps} shows our 15 GHz VLBA image of the jet of M87
in February of 2003.  No linear polarization is detected, but strong
circular polarization, $m_c = -0.49\pm0.10$\% is seen at the peak of 
the core. 

M87 was studied at the VLA in April of 2000 by \citet{BFM02}, who 
found no significant circular polarization at 8 GHz, 
$m_c = 0.01\pm0.10$; however, they did find linear polarization 
at the $1.7$\% level.  These results are the opposite of our finding
of strong circular polarization and no linear polarization.  It is
important to consider that their VLA results include contributions 
from the jet well beyond the region imaged by the VLBA, and the 
linear polarization they observe must come from beyond the
depolarizing screen which covers most of the VLBA scale jet 
\citep{ZT02}.  The fractional circular polarization they
measure may be diluted by emission beyond the VLBA scale, or the 
circular polarization may simply vary with frequency or time.  
Regardless, the differing linear polarizations observed between the
VLBA and VLA scale images underscores our point from \S{\ref{s:cores}}
that at least some of the lack of correlation between linear
and circular polarization for jet cores could be due to the two
quantities being produced in different regions of the unresolved
jet core.

M87 has been detected intermittently in integrated circular
polarization (with both signs) at 5 GHz and below in the 
historical observations cataloged by WdP83 and K84.
  
\begin{figure}
\begin{center}
\includegraphics[width=2in]{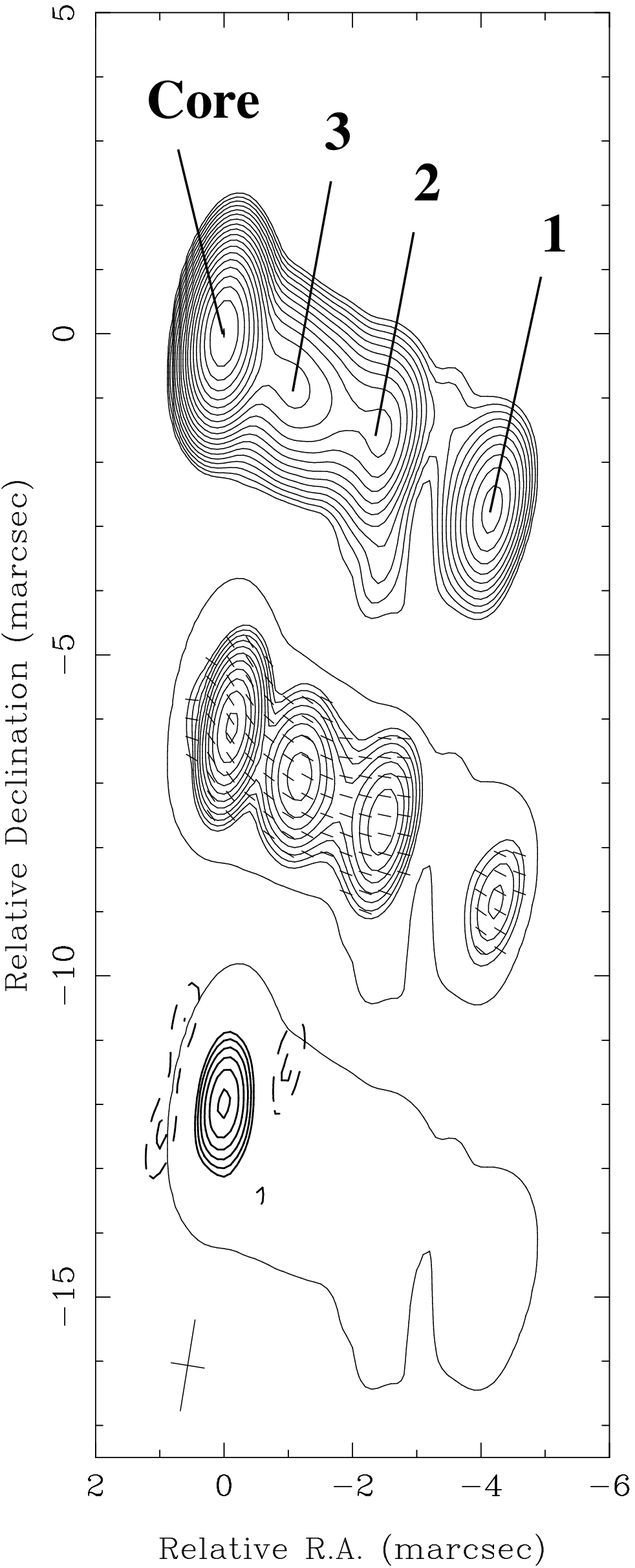}
\end{center}
\figcaption{\label{f:3c279cp}
Stokes-$I$ contours (top), linear polarization (middle), and circular polarization (bottom)
images of the inner 5 milli-arcseconds of the jet of 3C\,279 in November of 2002.  
A single Stokes-$I$ contour appears around the linear and circular polarization images
to show registration.  The Stokes-$I$ and linear polarization contours begin at $20$ mJy beam$^{-1}$,
and the circular polarization contours begin at $\pm5$ mJy beam$^{-1}$.  All contours increase in
steps of $\times\sqrt{2}$.  Negative circular polarization is indicated by dashed contours. 
The FWHM dimensions of the common
restoring beam is indicated by a cross in the lower left corner of the image.
Electric vector position angles (EVPA) for the linear polarization are 
indicated by tick marks in the middle image.  The locations of circular polarization 
measurements are indicated on the Stokes-$I$ map.
}
\end{figure}

\paragraph{1253$-$055 (3C\,279):}
Figure \ref{f:3c279cp} shows the first 5 milli-arcseconds of the jet of 3C\,279
at 15 GHz in November of 2002.  The top panel is Stokes I, the middle panel is
linear polarization, and the bottom panel is circular polarization.  
Positions where circular polarization measurements or limits were taken
are marked and numbered on the Stokes I map.  We detect strong circular
polarization of $m_c=+0.30\pm0.08$\% only on the jet core.  The sign and magnitude
of the circular polarization is consistent with five epochs of CP detections 
by HW99 at 15 GHz during 1996.  At 5 GHz, \citet{HAW01} did not detect circular
polarization with a limit of $|m_c| < 0.11$\%.  Historically, 3C\,279 has 
repeatedly and consistently shown a positive sign for its circular 
polarization in integrated measurements at 8 GHz and below during the 1970s
and early 80s (WdP83 and K84); however, \citet{AAP03} found $m_c = -0.17\pm0.01$\%
at 4.8 GHz with the UMRAO during 2001-2002.  This change in sign at the lower
frequency is particularly interesting as no change has yet occurred at 15 GHz.

\paragraph{1308$+$326:}
We find a limit of $|m_c| < 0.21$\% in the core at 15 GHz in November of 2003.
HW99 also found limits between 0.1 and 0.2\%  in the core at 15 GHz throughout
1996.

\paragraph{1510$-$089:}
We measure $m_c = +0.20\pm0.09$ \% in the core at 15 GHz in November of 
2002.  HW99 did not detect circular polarization in the core at 15 GHz
during 1996: three epochs had limits of 0.1 to 0.2\%, and 
for two epochs they reported peak CP values of $+0.2\pm0.2$ \%.  \citet{HAW01} find a 
limit of $|m_c| < 0.24$\% in the core at 5 GHz in December of 1996.

\paragraph{1641$+$399 (3C\,345):}
We find a limit of $|m_c| < 0.19$\% in the core at 15 GHz in November of
2002.  \citet{HW03} detected circular polarization in the core
at 15 GHz, $m_c = -0.38\pm012$\%, in January of 1998.  The same location
was optically thick at 8 GHz where they did not detect any circular
polarization in simultaneous measurements.  3C\,345 was detected
intermittently (and with both signs) at 8 GHz and below in the historical 
integrated measurement compiled by WdP83.

\paragraph{1730$-$130:}
We find a limit of $|m_c| < 0.17$\% in the core at 15 GHz in October of
2002.  The historical integrated measurements (WdP83 and K84) show 
several intermittent detections of positive circular polarization
at 5 GHz and below.

\paragraph{1749$+$096:}
We find a limit of $|m_c| < 0.14$\% in the core at 15 GHz in May of 2002.
HW99 also found typical limits of 0.2\%  in the core at 15 GHz throughout
1996.

\paragraph{1928$+$738:}
We measure $m_c = +0.19\pm0.09$ \% in the core at 15 GHz in June of 
2002.  HW99 also measured circular polarization in the core 15 GHz at
the limit of their sensitivity.  They found $m_c =+0.3\pm0.1$ \% in
two epochs, $m_c =+0.2\pm0.1$ \% in one epoch, and limits of 0.1 
and 0.2\% in their other two epochs during 1996.

\begin{figure}
\begin{center}
\includegraphics[width=2in]{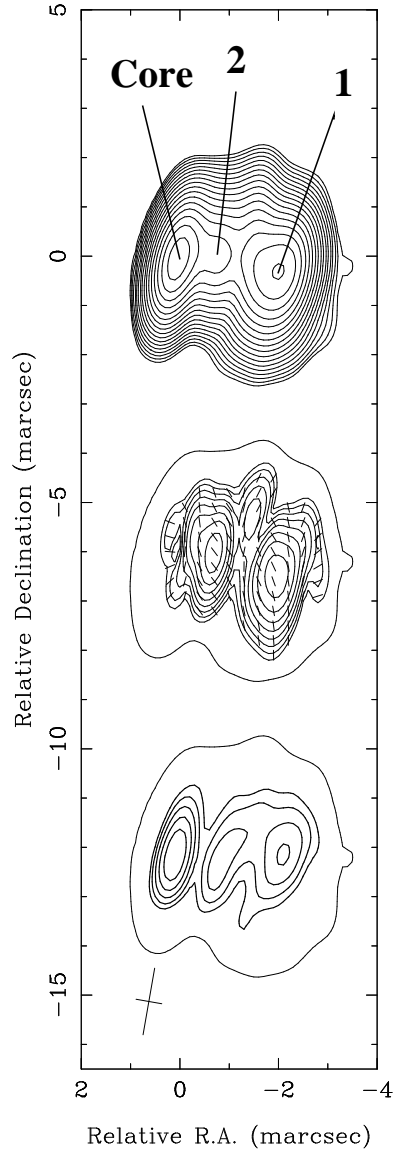}
\end{center}
\figcaption{\label{f:2134cp}
Stokes-$I$ contours (top), linear polarization (middle), and circular polarization (bottom)
images of 2134$+$004 in May of 2002.  
A single Stokes-$I$ contour appears around the linear and circular polarization images
to show registration.  The Stokes-$I$ and linear polarization contours begin at $5$ mJy beam$^{-1}$,
and the circular polarization contours begin at $\pm1$ mJy beam$^{-1}$.  All contours increase in
steps of $\times\sqrt{2}$.  Negative circular polarization is indicated by dashed contours. 
The FWHM dimensions of the common
restoring beam is indicated by a cross in the lower left corner of the image.
Electric vector position angles (EVPA) for the linear polarization are 
indicated by tick marks in the middle image.  The locations of circular polarization 
measurements are indicated on the Stokes-$I$ map.
}
\end{figure}

\paragraph{2134$+$004:}
Figure \ref{f:2134cp} shows our observations of 2134$+$004 at 15 GHz in May of 2002.  
The top panel is Stokes I, the middle panel is linear polarization, and the 
bottom panel is circular polarization.  Positions where circular polarization 
measurements or limits were taken are marked and numbered on the Stokes I map.  
We detect strong circular polarization of $m_c=+0.35\pm0.12$\% on the jet core,
we measure a similar value at location number 1 of $m_c=+0.29\pm0.12$\%, and
we find a limit on the CP at location 2 to be $|m_c| < 0.34$\%.

In recent integrated measurements from 2001--2002, \citet{AAP03} found strong
circular polarization at 4.8 GHz with $m_c = +0.62\pm0.06$\%.  2134$+$004 was
also detected repeatedly in the historical integrated measurements at 5 GHz
and below, always with positive circular polarization (WdP83 and K84).

\paragraph{2145$+$067:}
We find a limit of $|m_c| < 0.26$\% in the core at 15 GHz in March of
2003.  In recent integrated measurement from 2001--2002, \citet{AAP03} 
measured $m_c = -0.50\pm0.07$\% at 4.8 GHz with the UMRAO.
The historical integrated measurements (WdP83 and K84) show 
consistent, repeated detections of negative circular polarization
at 5 GHz and below.  Given this historical record of detection, it is
interesting that the 'limit' reported here came from measurements that
had a significance of $1.98\sigma$, just barely less than the 
$2\sigma$ cutoff we had for defining measurements versus limits.  
If we allow $1.98\sigma$ as a measurement, then our fractional level
for the core is $m_c = -0.17\pm0.09$ \%.

\paragraph{2200$+$420:}
We find a limit of $|m_c| < 0.21$\% in the core at 15 GHz in June of 2002.
WdP83 reported intermittent detections of both signs of circular polarization
at frequencies below 5 GHz in the historical integrated measurements.

\paragraph{2230$+$114 (CTA102):}
We find a limit of $|m_c| < 0.19$\% in the core at 15 GHz in October of 2002. 
\citet{HAW01} measured $m_c = -0.09\pm0.04$\% in the core at 5 GHz in December
of 1996.  CTA102 was intermittently detected (with both signs of CP) in the
historical integrated measurements (WdP83; K84) at 5 GHz and below.

\begin{figure}
\begin{center}
\includegraphics[width=2in]{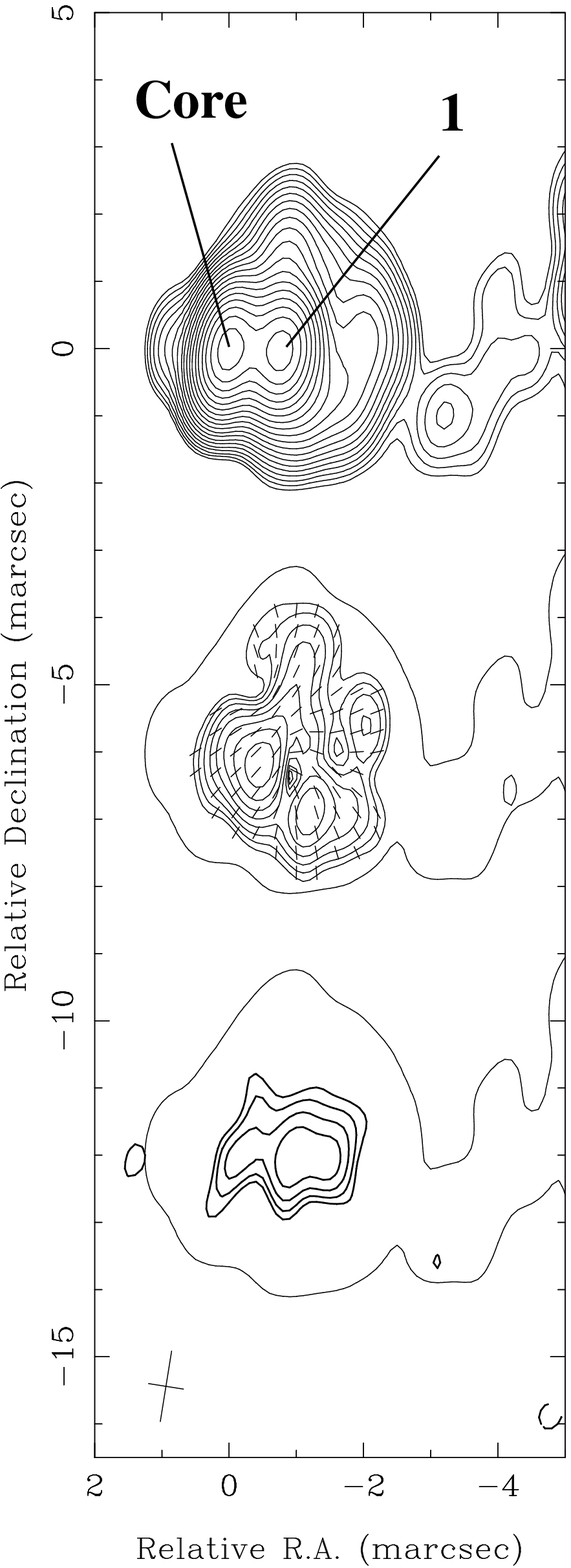}
\end{center}
\figcaption{\label{f:2251cp}
Stokes-$I$ contours (top), linear polarization (middle), and circular polarization (bottom)
images of the inner 4 milli-arcseconds of the jet of 3C\,454.3 in March of 2003.  
A single Stokes-$I$ contour appears around the linear and circular polarization images
to show registration.  The Stokes-$I$ and linear polarization contours begin at $5$ mJy beam$^{-1}$,
and the circular polarization contours begin at $\pm3$ mJy beam$^{-1}$.  All contours increase in
steps of $\times\sqrt{2}$.  Negative circular polarization is indicated by dashed contours. 
The FWHM dimensions of the common
restoring beam is indicated by a cross in the lower left corner of the image.
Electric vector position angles (EVPA) for the linear polarization are 
indicated by tick marks in the middle image.  The locations of circular polarization 
measurements are indicated on the Stokes-$I$ map.
}
\end{figure}

\paragraph{2251$+$158 (3C\,454.3):}
Figure \ref{f:2251cp} shows the first 4 milli-arcseconds of the jet of 
3C\,454.3 at 15 GHz in March of 2003.  Positions where circular polarization 
measurements or limits were taken are marked and numbered on the 
Stokes I map.  The first jet component (position 1) appears slightly
more strongly circularly polarized, $m_c = +0.38\pm0.10$\%, than
the core, $m_c = +0.23\pm0.10$\%; however, an alternative calibration
assuming no circular polarization for calibrating the phases (see
\S{\ref{s:phase}})  spreads the circular polarization between the core
and component 1 more evenly with both having $m_c \simeq 0.3$\%.

\citet{HAW01} measured $m_c = +0.08\pm0.04$\% in the core of 3C\,454.3 at 5 GHz 
in December of 1996, and the source was repeatedly detected with positive
circular polarization at 8 GHz and below in the integrated historical
measurements compiled by WdP83 and K84.

\paragraph{2345$-$167:}
We detect $m_c = +0.40\pm0.11$\% in the core at 15 GHz in October of 2002.
\citet{HAW01} found a limit of $|m_c| < 0.11$\% in the core at 5 GHz in December of
1996.  WdP83 and K84 reported intermittent detections (but always with positive 
sign) of circular polarization in integrated measurements at frequencies of 5 GHz
and below.

\section{Conclusions}
\label{s:conclude}

In this paper, we report first-epoch circular polarization results for
each of the 133 AGN in the MOJAVE program to monitor the VLBA structure and 
polarization of a flux limited sample of AGN jets at 15 GHz.  We found strong
circular polarization ($\geq 0.3$\%) in approximately 15\% of our sample.  The 
circular polarization we detected was usually associated with jet cores, typically 
in the range of $0.3-0.5$\% of the local Stokes-$I$.
Our ability to detect circular polarization on jet components was limited
by their relatively weak flux density; however, we did find a few strong jet 
components to be circularly polarized.  

We found no significant correlations between fractional circular polarization
of jet cores and optical source type, redshift, EGRET detections, linear polarization, 
or a number of other pc-scale jet properties listed in Table 4.  While there was no
clear relation between linear and circular polarization of jet cores, quasars
and BL Lac objects consistently had more linear than circular polarization, and 
the two galaxies with detected circular polarization had much more circular
than linear polarization.  We suggested that higher redshift sources may have
smaller $m_c/m_l$ ratios, not because they produce circular polarization
differently from low redshift sources, but rather because either (1) they
have less depolarization, and/or (2) we study them with less effective
linear resolution and therefore polarization from more of the jet
contributes to what we see as the ``core''.  

The lack of clear correlations between circular polarization and other
source properties is intriguing, particularly given that some sources, 
3C\,273 and 3C\,279 in particular, show consistently high levels of
circular polarization over years and decades of observation, even though 
they are both known to be highly variable with regular ($\sim$ annual)
ejections of new jet components.  It seems highly unlikely that these
sources would be circularly polarized in a very similar fashion from 
epoch to epoch due only to stochastic factors (such as a chance favorable 
magnetic field configuration).  It may be that we simply detect too few sources in 
circular polarization to find subtle trends between circular polarization 
and other source properties, or perhaps, even with the VLBA we are
not resolving the core region well enough to see where the circular 
polarization is being generated.

In addition to looking for overall trends in the circular polarization of
our sample, we also examined the structure of a few sources in detail.
3C\,84 was particularly interesting, showing the strongest circular
polarization in our sample, up to $2$\% in one location.  Our image 
revealed essentially the same structure as \citet{HW04} found in data 
taken five years earlier; however, a new circularly polarized spot 
appeared in our data, and the older spots of circular polarization, 
while maintaining the same relative positions, appear to have moved outward 
in the jet at a rate of approximately 0.06$c$.  3C\,273 also showed multiple 
locations of circular polarization in its jet, and we were able to plot circular 
polarization measurements for several strong jet components.  This plot revealed 
that, in 3C\,273, fractional circular polarization gets weaker at greater 
distances from the jet core.  Multi-frequency observations will be necessary 
for both sources so we can use opacity measurements to help understand the 
trends of circular polarization with jet position. 

\acknowledgments

The authors wish to acknowledge the contributions of the other members of the 
MOJAVE team: Hugh and Margo Aller, Tigran Arshakian, Marshall Cohen, Matthias Kadler, 
Ken Kellermann, Yuri Kovalev, Andrei Lobanov, Eduardo Ros, Rene Vermeulen, and Tony Zensus.
This work has been supported by Denison University, the National Radio Astronomy 
Observatory, and NSF grant 0406923-AST.  This research was also supported by 
an award from Research Corporation.



\begin{deluxetable}{cllcrrrrcc}
\tablecolumns{10}
\tablewidth{0pc}
\tabletypesize{\scriptsize}
\tablenum{1}
\tablecaption{Circular Polarization at Stokes $I$ Peak.\label{t:cp_tab1}}
\tablehead{\colhead{IAU} &&& \colhead{Opt.} && \colhead{$I_{peak}$} & 
\colhead{$m_l$} & \colhead{$V_{peak}$} & \colhead{$m_c$} &\\
\colhead{Name} & \colhead{Alias} & \colhead{z} &
\colhead{Class} & \colhead{Epoch} & \colhead{(Jy beam$^{-1}$)} & \colhead{(\%)} & \colhead{(mJy beam$^{-1}$)} &
\colhead{(\%)} & \colhead{$\sigma$} \\
\colhead{(1)} & \colhead{(2)} & \colhead{(3)} & \colhead{(4)} &
\colhead{(5)} & \colhead{(6)} & \colhead{(7)} & \colhead{(8)} &
\colhead{(9)} & \colhead{(10)}}
\startdata
$0003-066$ & NRAO 005 & $0.347$ & B & 2003 Feb 06 & $1.83$ & $4.30$ & $+1.55\pm1.82$ & $<0.20$ & \nodata\\
$0007+106$ & III Zw 2 & $0.0893$ & G & 2004 Feb 11 & $1.35$ & $< 0.08$ & $-2.74\pm1.54$ & $<0.32$ & \nodata\\
$0016+731$ & \nodata & $1.781$ & Q & 2003 Aug 28 & $0.99$ & $0.97$ & $+1.02\pm1.73$ & $<0.35$ & \nodata\\
$0048-097$ & \nodata & \nodata & B & 2003 Aug 28 & $0.70$ & $3.57$ & $-0.74\pm1.78$ & $<0.51$ & \nodata\\
$0059+581$ & \nodata & \nodata & U & 2002 Nov 23 & $3.16$ & $1.96$ & $+1.22\pm2.68$ & $<0.17$ & \nodata\\
$0106+013$ & \nodata & $2.107$ & Q & 2003 Mar 29 & $2.13$ & $1.54$ & $+1.27\pm2.07$ & $<0.19$ & \nodata\\
$0109+224$ & \nodata & \nodata & B & 2002 Jun 15 & $0.93$ & $4.58$ & $-0.82\pm1.14$ & $<0.25$ & \nodata\\
$0119+115$ & \nodata & $0.57$ & Q & 2003 Mar 01 & $0.88$ & $5.58$ & $-0.01\pm1.18$ & $<0.27$ & \nodata\\
$0133+476$ & DA 55   & $0.859$ & Q & 2002 Jun 15 & $4.20$ & $3.14$ & $-7.79\pm3.69$ & $-0.18\pm0.09$ & 2.1\\
$0202+149$ & 4C $+$15.05 & $0.405$ & Q & 2002 Oct 09 & $1.23$ & $0.08$ & $+0.23\pm1.39$ & $<0.23$ & \nodata\\
$0202+319$ & \nodata & $1.466$ & Q & 2003 Mar 29 & $2.10$ & $1.57$ & $+2.03\pm2.00$ & $<0.19$ & \nodata\\
$0212+735$ & \nodata & $2.367$ & Q & 2003 May 09 & $1.25$ & $10.90$ & $+2.46\pm1.35$ & $<0.31$ & \nodata\\
$0215+015$ & \nodata & $1.715$ & B & 2002 May 31 & $0.75$ & $1.39$ & $+2.10\pm0.89$ & $+0.28\pm0.12$ & 2.4\\
$0224+671$ & 4C $+$67.05 & \nodata & U & 2002 Nov 23 & $0.68$ & $0.86$ & $0.00\pm0.86$ & $<0.25$ & \nodata\\
$0234+285$ & CTD 20 & $1.213$ & Q & 2002 Nov 23 & $3.42$ & $0.79$ & $+2.82\pm2.88$ & $<0.17$ & \nodata\\
$0235+164$ & \nodata & $0.94$ & B & 2003 Mar 01 & $1.45$ & $4.85$ & $-1.45\pm1.45$ & $<0.20$ & \nodata\\
$0238-084$ & NGC 1052 & $0.0049$ & G & 2002 Oct 09 & $0.34$ & $< 0.58$ & $+0.83\pm2.43$ & $<1.44$ & \nodata\\
$0300+470$ & 4C $+$47.08 & \nodata & B & 2002 Nov 23 & $1.04$ & $1.02$ & $+0.12\pm1.09$ & $<0.21$ & \nodata\\
$0316+413$ & 3C 84 & $0.01756$ & G & 2003 Mar 01 & $2.65$ & $< 0.04$ & $-4.87\pm3.67$ & $<0.32$ & \nodata\\
$0333+321$ & NRAO 140 & $1.263$ & Q & 2003 Mar 29 & $1.73$ & $0.49$ & $-4.88\pm1.82$ & $-0.28\pm0.10$ & 2.7\\
$0336-019$ & CTA 26 & $0.852$ & Q & 2003 Feb 06 & $2.72$ & $4.37$ & $+1.47\pm2.47$ & $<0.18$ & \nodata\\
$0403-132$ & \nodata & $0.571$ & Q & 2002 May 31 & $1.32$ & $0.62$ & $+0.21\pm1.21$ & $<0.18$ & \nodata\\
$0415+379$ & 3C 111 & $0.0491$ & G & 2002 Oct 09 & $1.17$ & $< 0.08$ & $-0.22\pm1.21$ & $<0.21$ & \nodata\\
$0420-014$ & \nodata & $0.915$ & Q & 2003 Mar 01 & $9.46$ & $0.87$ & $+1.58\pm7.62$ & $<0.16$ & \nodata\\
$0422+004$ & \nodata & \nodata & B & 2002 Jun 15 & $1.64$ & $8.11$ & $+1.18\pm1.58$ & $<0.19$ & \nodata\\
$0430+052$ & 3C 120 & $0.033$ & G & 2003 Feb 06 & $1.04$ & $< 0.09$ & $+1.92\pm1.10$ & $<0.29$ & \nodata\\
$0446+112$ & \nodata & $1.207$ & Q & 2002 May 31 & $2.11$ & $1.14$ & $-2.30\pm1.85$ & $<0.20$ & \nodata\\
$0458-020$ & \nodata & $2.286$ & Q & 2003 Mar 01 & $1.04$ & $1.29$ & $-0.03\pm1.23$ & $<0.24$ & \nodata\\
$0528+134$ & \nodata & $2.07$ & Q & 2003 Feb 06 & $2.64$ & $4.55$ & $+0.95\pm2.37$ & $<0.18$ & \nodata\\
$0529+075$ & \nodata & \nodata & U & 2002 May 31 & $0.69$ & $0.41$ & $-2.17\pm2.46$ & $<0.71$ & \nodata\\
$0529+483$ & \nodata & $1.162$ & Q & 2002 Oct 09 & $0.90$ & $1.99$ & $-0.22\pm0.95$ & $<0.21$ & \nodata\\
$0552+398$ & DA 193 & $2.363$ & Q & 2003 Mar 29 & $2.39$ & $1.60$ & $-2.48\pm2.23$ & $<0.20$ & \nodata\\
$0605-085$ & \nodata & $0.872$ & Q & 2003 Feb 06 & $1.00$ & $2.54$ & $-0.92\pm1.32$ & $<0.26$ & \nodata\\
$0607-157$ & \nodata & $0.324$ & Q & 2003 Mar 01 & $4.04$ & $3.65$ & $+5.09\pm3.42$ & $<0.21$ & \nodata\\
$0642+449$ & OH 471 & $3.408$ & Q & 2003 Mar 29 & $3.46$ & $2.94$ & $+13.57\pm2.96$ & $+0.39\pm0.09$ & 4.6\\
$0648-165$ & \nodata & \nodata & U & 2002 Nov 23 & $2.31$ & $0.99$ & $-1.93\pm1.97$ & $<0.17$ & \nodata\\
$0716+714$ & \nodata & \nodata & B & 2003 Aug 28 & $2.51$ & $4.84$ & $+9.34\pm2.82$ & $+0.37\pm0.11$ & 3.3\\
$0727-115$ & \nodata & $1.591$ & Q & 2003 Mar 01 & $4.09$ & $3.14$ & $-3.44\pm3.68$ & $<0.18$ & \nodata\\
$0730+504$ & \nodata & $0.72$ & Q & 2002 Jun 15 & $0.72$ & $0.65$ & $-3.78\pm1.20$ & $-0.53\pm0.17$ & 3.2\\
$0735+178$ & \nodata & \nodata & B & 2002 Nov 23 & $0.85$ & $3.36$ & $-2.57\pm0.92$ & $-0.30\pm0.11$ & 2.8\\
$0736+017$ & \nodata & $0.191$ & Q & 2003 Mar 01 & $1.49$ & $0.92$ & $+6.66\pm1.41$ & $+0.45\pm0.09$ & 4.7\\
$0738+313$ & OI 363 & $0.63$ & Q & 2003 Feb 06 & $0.51$ & $0.42$ & $-1.79\pm1.65$ & $<0.68$ & \nodata\\
$0742+103$ & \nodata & \nodata & U & 2004 Feb 11 & $0.59$ & $< 0.16$ & $-0.31\pm0.96$ & $<0.33$ & \nodata\\
$0748+126$ & \nodata & $0.889$ & Q & 2003 Feb 06 & $1.79$ & $3.37$ & $+2.69\pm1.66$ & $<0.24$ & \nodata\\
$0754+100$ & \nodata & $0.266$ & B & 2002 Nov 23 & $1.38$ & $6.12$ & $-2.46\pm1.36$ & $<0.28$ & \nodata\\
$0804+499$ & \nodata & $1.432$ & Q & 2004 Feb 11 & $0.61$ & $0.60$ & $-0.27\pm1.11$ & $<0.36$ & \nodata\\
$0805-077$ & \nodata & $1.837$ & Q & 2002 Jun 15 & $1.28$ & $0.64$ & $+0.71\pm1.41$ & $<0.22$ & \nodata\\
$0808+019$ & \nodata & \nodata & B & 2004 Feb 11 & $0.46$ & $8.75$ & $+0.43\pm0.92$ & $<0.41$ & \nodata\\
$0814+425$ & \nodata & \nodata & B & 2004 Feb 11 & $0.67$ & $1.79$ & $+0.47\pm0.96$ & $<0.28$ & \nodata\\
$0823+033$ & \nodata & $0.506$ & B & 2003 Jun 15 & $0.97$ & $1.06$ & $+0.20\pm1.23$ & $<0.26$ & \nodata\\
$0827+243$ & \nodata & $0.941$ & Q & 2002 May 31 & $1.77$ & $0.08$ & $-1.37\pm1.64$ & $<0.19$ & \nodata\\
$0829+046$ & \nodata & $0.18$ & B & 2003 Aug 28 & $0.43$ & $0.78$ & $+0.60\pm1.69$ & $<0.80$ & \nodata\\
$0836+710$ & \nodata & $2.218$ & Q & 2003 Mar 29 & $1.19$ & $0.70$ & $+4.72\pm1.20$ & $+0.40\pm0.10$ & 3.9\\
$0851+202$ & OJ 287 & $0.306$ & B & 2002 Oct 09 & $3.48$ & $1.52$ & $-6.90\pm2.88$ & $-0.20\pm0.08$ & 2.4\\
$0906+015$ & 4C $+$01.24 & $1.018$ & Q & 2003 Mar 29 & $1.91$ & $5.59$ & $-0.23\pm1.80$ & $<0.19$ & \nodata\\
$0917+624$ & \nodata & $1.446$ & Q & 2002 Jun 15 & $0.57$ & $3.61$ & $-1.01\pm1.27$ & $<0.44$ & \nodata\\
$0923+392$ & 4C $+$39.25 & $0.698$ & Q & 2002 Oct 09 & $4.33$ & $2.05$ & $+4.09\pm3.76$ & $<0.18$ & \nodata\\
$0945+408$ & \nodata & $1.252$ & Q & 2002 Oct 09 & $1.00$ & $1.84$ & $+3.03\pm1.02$ & $+0.30\pm0.10$ & 3.0\\
$0955+476$ & \nodata & $1.873$ & Q & 2002 Jun 15 & $1.56$ & $0.86$ & $-2.73\pm1.60$ & $<0.28$ & \nodata\\
$1036+054$ & \nodata & \nodata & U & 2002 May 31 & $2.55$ & $1.45$ & $+1.86\pm1.99$ & $<0.16$ & \nodata\\
$1038+064$ & \nodata & $1.265$ & Q & 2002 May 31 & $1.03$ & $0.60$ & $-1.97\pm1.06$ & $<0.29$ & \nodata\\
$1045-188$ & \nodata & $0.595$ & Q & 2002 Jun 15 & $1.11$ & $3.26$ & $+1.99\pm1.64$ & $<0.33$ & \nodata\\
$1055+018$ & 4C $+$01.28 & $0.888$ & Q & 2003 Feb 06 & $3.15$ & $3.66$ & $+10.10\pm2.82$ & $+0.32\pm0.09$ & 3.6\\
$1124-186$ & \nodata & $1.048$ & Q & 2003 Mar 01 & $1.30$ & $1.94$ & $+2.67\pm1.49$ & $<0.32$ & \nodata\\
$1127-145$ & \nodata & $1.187$ & Q & 2002 Oct 09 & $1.07$ & $5.49$ & $-2.60\pm1.12$ & $-0.24\pm0.10$ & 2.3\\
$1150+812$ & \nodata & $1.25$ & Q & 2002 Jun 15 & $0.98$ & $0.72$ & $+1.23\pm1.20$ & $<0.25$ & \nodata\\
$1156+295$ & 4C $+$29.45 & $0.729$ & Q & 2002 Nov 23 & $1.86$ & $2.43$ & $-5.02\pm1.70$ & $-0.27\pm0.09$ & 3.0\\
$1213-172$ & \nodata & \nodata & U & 2002 Jun 15 & $1.11$ & $4.18$ & $+0.61\pm1.39$ & $<0.25$ & \nodata\\
$1219+044$ & \nodata & $0.965$ & Q & 2002 Jun 15 & $0.91$ & $0.22$ & $+1.72\pm1.12$ & $<0.31$ & \nodata\\
$1222+216$ & \nodata & $0.435$ & Q & 2002 May 31 & $0.56$ & $4.15$ & $-1.13\pm0.81$ & $<0.35$ & \nodata\\
$1226+023$ & 3C 273 & $0.158$ & Q & 2002 Oct 09 & $8.99$ & $3.74$ & $-40.28\pm8.07$ & $-0.45\pm0.09$ & 5.0\\
$1228+126$ & M87 & $0.0044$ & G & 2003 Feb 06 & $1.28$ & $< 0.07$ & $-6.32\pm1.34$ & $-0.49\pm0.10$ & 4.7\\
$1253-055$ & 3C 279 & $0.538$ & Q & 2002 Nov 23 & $10.31$ & $5.53$ & $+30.52\pm8.67$ & $+0.30\pm0.08$ & 3.5\\
$1308+326$ & \nodata & $0.997$ & Q & 2002 Nov 23 & $2.00$ & $3.26$ & $+2.30\pm1.99$ & $<0.21$ & \nodata\\
$1324+224$ & \nodata & $1.4$ & Q & 2002 Oct 09 & $0.55$ & $1.33$ & $+1.10\pm0.96$ & $<0.37$ & \nodata\\
$1334-127$ & \nodata & $0.539$ & Q & 2003 Mar 01 & $4.31$ & $1.43$ & $+12.71\pm4.47$ & $+0.29\pm0.10$ & 2.8\\
$1413+135$ & \nodata & $0.247$ & B & 2002 Nov 23 & $0.88$ & $< 0.10$ & $-1.89\pm1.10$ & $<0.34$ & \nodata\\
$1417+385$ & \nodata & $1.832$ & Q & 2002 Jun 15 & $0.84$ & $2.04$ & $-0.51\pm1.08$ & $<0.26$ & \nodata\\
$1458+718$ & 3C 309.1 & $0.904$ & Q & 2003 Aug 28 & $0.58$ & $1.43$ & $+0.77\pm1.39$ & $<0.48$ & \nodata\\
$1502+106$ & 4C $+$10.39 & $1.833$ & Q & 2003 Mar 29 & $1.45$ & $1.01$ & $-1.54\pm1.40$ & $<0.20$ & \nodata\\
$1504-166$ & \nodata & $0.876$ & Q & 2003 Feb 06 & $0.93$ & $1.76$ & $+2.43\pm1.07$ & $+0.26\pm0.12$ & 2.3\\
$1510-089$ & \nodata & $0.36$ & Q & 2002 Nov 23 & $2.73$ & $3.88$ & $+5.51\pm2.40$ & $+0.20\pm0.09$ & 2.3\\
$1538+149$ & 4C $+$14.60 & $0.605$ & B & 2004 Feb 11 & $0.97$ & $11.50$ & $-2.80\pm1.82$ & $<0.47$ & \nodata\\
$1546+027$ & \nodata & $0.412$ & Q & 2003 Mar 29 & $1.42$ & $1.81$ & $-0.88\pm1.66$ & $<0.24$ & \nodata\\
$1548+056$ & 4C $+$05.64 & $1.422$ & Q & 2003 Mar 01 & $1.66$ & $3.24$ & $-0.38\pm1.68$ & $<0.20$ & \nodata\\
$1606+106$ & 4C $+$10.45 & $1.226$ & Q & 2002 Nov 23 & $1.48$ & $2.55$ & $-0.06\pm1.59$ & $<0.21$ & \nodata\\
$1611+343$ & DA 406 & $1.401$ & Q & 2003 Feb 06 & $2.72$ & $2.97$ & $-3.44\pm2.45$ & $<0.22$ & \nodata\\
$1633+382$ & 4C $+$38.41 & $1.807$ & Q & 2003 Mar 29 & $3.48$ & $1.58$ & $-13.59\pm3.02$ & $-0.39\pm0.09$ & 4.5\\
$1637+574$ & \nodata & $0.751$ & Q & 2002 May 31 & $1.65$ & $1.64$ & $+0.67\pm1.39$ & $<0.17$ & \nodata\\
$1638+398$ & NRAO 512 & $1.666$ & Q & 2003 Mar 01 & $0.68$ & $7.10$ & $-1.15\pm0.97$ & $<0.31$ & \nodata\\
$1641+399$ & 3C 345 & $0.594$ & Q & 2002 Nov 23 & $2.95$ & $3.53$ & $-0.04\pm2.83$ & $<0.19$ & \nodata\\
$1655+077$ & \nodata & $0.621$ & Q & 2002 Oct 09 & $0.86$ & $3.20$ & $+2.40\pm0.93$ & $+0.28\pm0.11$ & 2.6\\
$1726+455$ & \nodata & $0.714$ & Q & 2002 May 31 & $2.07$ & $3.71$ & $-0.64\pm1.65$ & $<0.16$ & \nodata\\
$1730-130$ & NRAO 530 & $0.902$ & Q & 2002 Oct 09 & $4.75$ & $0.34$ & $+0.37\pm3.92$ & $<0.17$ & \nodata\\
$1739+522$ & 4C $+$51.37 & $1.379$ & Q & 2003 Mar 29 & $1.30$ & $2.86$ & $-0.59\pm1.32$ & $<0.20$ & \nodata\\
$1741-038$ & \nodata & $1.057$ & Q & 2003 Mar 01 & $5.96$ & $0.65$ & $-7.59\pm5.34$ & $<0.22$ & \nodata\\
$1749+096$ & 4C $+$09.57 & $0.32$ & B & 2002 May 31 & $3.13$ & $1.19$ & $+1.10\pm2.28$ & $<0.14$ & \nodata\\
$1751+288$ & \nodata & \nodata & U & 2002 Jun 15 & $1.84$ & $0.72$ & $+1.62\pm1.75$ & $<0.19$ & \nodata\\
$1758+388$ & \nodata & $2.092$ & Q & 2003 May 09 & $1.26$ & $1.21$ & $-1.30\pm1.07$ & $<0.19$ & \nodata\\
$1800+440$ & \nodata & $0.663$ & Q & 2003 Mar 29 & $1.10$ & $1.32$ & $+7.71\pm1.21$ & $+0.70\pm0.11$ & 6.3\\
$1803+784$ & \nodata & $0.68$ & B & 2003 Feb 06 & $1.33$ & $2.50$ & $+1.27\pm1.35$ & $<0.20$ & \nodata\\
$1823+568$ & 4C $+$56.27 & $0.663$ & B & 2003 May 09 & $1.32$ & $8.05$ & $+2.96\pm1.17$ & $+0.22\pm0.09$ & 2.5\\
$1828+487$ & 3C 380 & $0.692$ & Q & 2003 Mar 29 & $1.28$ & $2.26$ & $-0.02\pm3.27$ & $<0.51$ & \nodata\\
$1849+670$ & \nodata & $0.657$ & Q & 2003 Jun 15 & $1.59$ & $3.34$ & $-4.10\pm1.84$ & $-0.26\pm0.12$ & 2.2\\
$1928+738$ & 4C $+$73.18 & $0.303$ & Q & 2002 Jun 15 & $2.47$ & $0.12$ & $+4.78\pm2.23$ & $+0.19\pm0.09$ & 2.1\\
$1936-155$ & \nodata & $1.657$ & Q & 2002 May 31 & $0.73$ & $1.59$ & $+2.65\pm0.88$ & $+0.36\pm0.12$ & 3.0\\
$1957+405$ & Cygnus A & $0.056$ & G & 2002 Nov 23 & $0.64$ & $< 0.17$ & $+1.21\pm1.85$ & $<0.58$ & \nodata\\
$1958-179$ & \nodata & $0.652$ & Q & 2002 May 31 & $0.95$ & $0.48$ & $+2.56\pm1.03$ & $+0.27\pm0.11$ & 2.5\\
$2005+403$ & \nodata & $1.736$ & Q & 2003 Mar 01 & $1.56$ & $3.94$ & $+0.94\pm1.65$ & $<0.21$ & \nodata\\
$2008-159$ & \nodata & $1.18$ & Q & 2003 Jun 15 & $1.96$ & $< 0.05$ & $+2.80\pm2.06$ & $<0.25$ & \nodata\\
$2021+317$ & 4C $+$31.56 & \nodata & U & 2003 Feb 06 & $0.39$ & $3.22$ & $-0.04\pm0.85$ & $<0.43$ & \nodata\\
$2021+614$ & \nodata & $0.227$ & G & 2003 Mar 01 & $0.99$ & $< 0.11$ & $+0.18\pm1.18$ & $<0.24$ & \nodata\\
$2037+511$ & 3C 418 & $1.687$ & Q & 2002 Oct 09 & $1.62$ & $0.97$ & $+1.63\pm1.54$ & $<0.20$ & \nodata\\
$2121+053$ & \nodata & $1.941$ & Q & 2003 Mar 29 & $2.13$ & $6.56$ & $-1.60\pm1.92$ & $<0.18$ & \nodata\\
$2128-123$ & \nodata & $0.501$ & Q & 2003 May 09 & $1.78$ & $1.72$ & $+3.86\pm1.63$ & $+0.22\pm0.09$ & 2.4\\
$2131-021$ & 4C $-$02.81 & $1.285$ & B & 2003 May 09 & $1.14$ & $5.20$ & $+1.11\pm1.10$ & $<0.19$ & \nodata\\
$2134+004$ & \nodata & $1.932$ & Q & 2002 May 31 & $1.62$ & $0.38$ & $+5.63\pm1.89$ & $+0.35\pm0.12$ & 3.0\\
$2136+141$ & OX 161 & $2.427$ & Q & 2002 Nov 23 & $2.15$ & $3.02$ & $+9.45\pm2.23$ & $+0.44\pm0.10$ & 4.2\\
$2145+067$ & 4C $+$06.69 & $0.999$ & Q & 2003 Mar 01 & $5.88$ & $1.56$ & $-10.11\pm5.12$ & $<0.26$ & \nodata\\
$2155-152$ & \nodata & $0.672$ & Q & 2003 Mar 29 & $1.27$ & $3.66$ & $-1.97\pm1.43$ & $<0.27$ & \nodata\\
$2200+420$ & BL Lac & $0.069$ & B & 2002 Jun 15 & $1.25$ & $2.65$ & $-0.15\pm1.30$ & $<0.21$ & \nodata\\
$2201+171$ & \nodata & $1.076$ & Q & 2003 Jun 15 & $1.83$ & $1.90$ & $+3.42\pm1.70$ & $+0.19\pm0.09$ & 2.0\\
$2201+315$ & 4C $+$31.63 & $0.298$ & Q & 2002 Nov 23 & $2.23$ & $2.82$ & $+0.74\pm2.18$ & $<0.20$ & \nodata\\
$2209+236$ & \nodata & $1.125$ & Q & 2003 Mar 01 & $0.62$ & $10.08$ & $-1.98\pm1.06$ & $<0.49$ & \nodata\\
$2216-038$ & \nodata & $0.901$ & Q & 2002 May 31 & $1.95$ & $2.17$ & $+0.27\pm1.64$ & $<0.17$ & \nodata\\
$2223-052$ & 3C 446 & $1.404$ & Q & 2002 Oct 09 & $3.72$ & $3.81$ & $+4.45\pm3.67$ & $<0.22$ & \nodata\\
$2227-088$ & PHL 5225 & $1.562$ & Q & 2003 Mar 29 & $1.88$ & $0.34$ & $+0.67\pm1.71$ & $<0.18$ & \nodata\\
$2230+114$ & CTA 102 & $1.037$ & Q & 2002 Oct 09 & $2.32$ & $1.67$ & $+0.12\pm2.22$ & $<0.19$ & \nodata\\
$2243-123$ & \nodata & $0.63$ & Q & 2002 Oct 09 & $1.03$ & $1.32$ & $+0.97\pm1.27$ & $<0.25$ & \nodata\\
$2251+158$ & 3C 454.3 & $0.859$ & Q & 2003 Mar 29 & $3.25$ & $1.05$ & $+7.35\pm3.32$ & $+0.23\pm0.10$ & 2.2\\
$2331+073$ & \nodata & \nodata & U & 2002 Jun 15 & $0.96$ & $3.38$ & $+1.26\pm1.06$ & $<0.24$ & \nodata\\
$2345-167$ & \nodata & $0.576$ & Q & 2002 Oct 09 & $1.39$ & $0.77$ & $+5.55\pm1.48$ & $+0.40\pm0.11$ & 3.8\\
$2351+456$ & 4C $+$45.51 & $1.986$ & Q & 2002 May 31 & $1.39$ & $2.09$ & $+2.83\pm1.45$ & $<0.31$ & \nodata\\
\enddata
\tablecomments{Columns are as follows: (1) IAU Name (B1950); (2) Other Name;  
(3) Redshift \citep[see][for references]{LH05};
(4) Optical classification where Q = quasar, B = BL Lac object, G = active galaxy,
and U = unidentified; (5) Epoch of first observation in MOJAVE program; (6) Stokes $I$ at the map peak; 
(7) Fractional linear polarization of the Stokes $I$ peak; (8) Maximum circular polarization measured within two pixels of
the Stokes $I$ peak; (9) Fractional circular polarization corresponding to Stokes-$V$ in column (8);
(10) Significance of the circular polarization measurement: $|m_c|/\Delta m_c$.}
\end{deluxetable}

\begin{deluxetable}{crrrrrcc}
\tablecolumns{8}
\tablewidth{0pc}
\tabletypesize{\scriptsize}
\tablenum{2}
\tablecaption{Circular Polarization on Off-Peak Jet Cores.\label{t:cp_tab2}}
\tablehead{\colhead{IAU} & \colhead{$R$} & \colhead{$\theta$} & \colhead{$I_{core}$} & 
\colhead{$m_l$} & \colhead{$V_{core}$} & \colhead{$m_c$} &\\
\colhead{Name} & \colhead{(marcsec)} & \colhead{(deg.)} &
\colhead{(Jy beam$^{-1}$)} & \colhead{(\%)} & \colhead{(mJy beam$^{-1}$)} &
\colhead{(\%)} & \colhead{$\sigma$} \\
\colhead{(1)} & \colhead{(2)} & \colhead{(3)} & \colhead{(4)} &
\colhead{(5)} & \colhead{(6)} & \colhead{(7)} & \colhead{(8)}}
\startdata
$0212+735$ & $0.67$ & $-63.4$ & $0.79$ & $5.97$ & $+0.21\pm0.90$ & $<0.23$ & \nodata\\
$0316+413$ & $0.28$ & $45.0$ & $2.26$ & $< 0.05$ & $-8.61\pm3.15$ & $-0.38\pm0.14$ & 2.7\\
$0738+313$ & $3.71$ & $-4.6$ & $0.33$ & $< 0.32$ & $+0.37\pm1.12$ & $<0.68$ & \nodata\\
$0923+392$ & $2.75$ & $-79.5$ & $0.19$ & $< 0.78$ & $-0.95\pm0.81$ & $<0.92$ & \nodata\\
$1213-172$ & $0.50$ & $-53.1$ & $0.53$ & $1.84$ & $+0.86\pm0.71$ & $<0.30$ & \nodata\\
$1226+023$ & $0.72$ & $56.3$ & $2.78$ & $0.62$ & $-18.77\pm2.71$ & $-0.67\pm0.10$ & 6.9\\
$1548+056$ & $0.63$ & $161.6$ & $1.25$ & $1.51$ & $-0.20\pm1.29$ & $<0.21$ & \nodata\\
$2128-123$ & $1.12$ & $26.6$ & $0.43$ & $< 0.20$ & $-0.20\pm0.51$ & $<0.23$ & \nodata\\
\enddata
\tablecomments{Columns are as follows: (1) IAU Name (B1950); (2) Radial position of the core relative to the map peak;  
(3) Structural position angle of the core relative to the map peak;
(4) Peak Stokes $I$ at the core position; 
(5) Fractional linear polarization at the core position; (6) Maximum circular polarization measured within two pixels of
the core; (7) Fractional circular polarization corresponding to Stokes $V$ in column(6);
(8) Significance of the circular polarization measurement:  $|m_c|/\Delta m_c$.}
\end{deluxetable}

\begin{deluxetable}{crrrrrcc}
\tablecolumns{8}
\tablewidth{0pc}
\tabletypesize{\scriptsize}
\tablenum{3}
\tablecaption{Circular Polarization on Strong Jet Components.\label{t:cp_tab3}}
\tablehead{\colhead{IAU} & \colhead{$R$} & \colhead{$\theta$} & \colhead{$I_{comp}$} & 
\colhead{$m_l$} & \colhead{$V_{comp}$} & \colhead{$m_c$} &\\
\colhead{Name} & \colhead{(marcsec)} & \colhead{(deg.)} &
\colhead{(Jy beam$^{-1}$)} & \colhead{(\%)} & \colhead{(mJy beam$^{-1}$)} &
\colhead{(\%)} & \colhead{$\sigma$} \\
\colhead{(1)} & \colhead{(2)} & \colhead{(3)} & \colhead{(4)} &
\colhead{(5)} & \colhead{(6)} & \colhead{(7)} & \colhead{(8)}}
\startdata
$0212+735$ & $0.67$ & $116.6$ & $1.25$ & $10.90$ & $+2.46\pm1.35$ & $<0.31$ & \nodata\\
$0224+671$ & $0.41$ & $14.0$ & $0.60$ & $1.17$ & $+0.12\pm0.78$ & $<0.26$ & \nodata\\
$0316+413$ & $1.00$ & $-143.1$ & $2.00$ & $< 0.06$ & $-8.43\pm2.82$ & $-0.42\pm0.14$ & 3.0\\
$0316+413$ & $1.77$ & $-137.3$ & $0.65$ & $< 0.17$ & $-8.93\pm1.13$ & $-1.37\pm0.17$ & 7.9\\
$0316+413$ & $1.94$ & $-168.1$ & $0.34$ & $< 0.34$ & $+7.56\pm0.84$ & $+2.25\pm0.25$ & 9.0\\
$0552+398$ & $0.85$ & $-69.4$ & $0.33$ & $< 0.37$ & $-0.34\pm0.48$ & $<0.29$ & \nodata\\
$0738+313$ & $3.71$ & $175.4$ & $0.51$ & $0.42$ & $-1.79\pm1.65$ & $<0.68$ & \nodata\\
$0754+100$ & $1.22$ & $9.5$ & $0.31$ & $4.03$ & $-0.13\pm0.39$ & $<0.25$ & \nodata\\
$0851+202$ & $1.25$ & $-118.6$ & $0.37$ & $7.73$ & $+0.61\pm0.52$ & $<0.30$ & \nodata\\
$0923+392$ & $2.75$ & $100.5$ & $4.33$ & $2.05$ & $+4.09\pm3.76$ & $<0.18$ & \nodata\\
$0945+408$ & $0.94$ & $122.0$ & $0.33$ & $2.77$ & $+0.75\pm0.44$ & $<0.36$ & \nodata\\
$1038+064$ & $0.71$ & $171.9$ & $0.49$ & $1.29$ & $-0.81\pm0.56$ & $<0.28$ & \nodata\\
$1127-145$ & $0.51$ & $101.3$ & $0.57$ & $7.30$ & $-0.35\pm0.63$ & $<0.22$ & \nodata\\
$1213-172$ & $0.50$ & $126.9$ & $1.11$ & $4.18$ & $+0.61\pm1.39$ & $<0.25$ & \nodata\\
$1226+023$ & $0.72$ & $-123.7$ & $8.99$ & $3.74$ & $-40.28\pm8.07$ & $-0.45\pm0.09$ & 5.0\\
$1226+023$ & $2.56$ & $-120.6$ & $0.46$ & $6.33$ & $-2.59\pm1.18$ & $-0.56\pm0.26$ & 2.2\\
$1226+023$ & $3.55$ & $-130.4$ & $1.28$ & $7.08$ & $-4.88\pm1.59$ & $-0.38\pm0.12$ & 3.1\\
$1226+023$ & $4.53$ & $-131.4$ & $0.82$ & $4.20$ & $-1.00\pm1.33$ & $<0.32$ & \nodata\\
$1226+023$ & $9.75$ & $-115.5$ & $1.49$ & $9.97$ & $+0.10\pm1.73$ & $<0.23$ & \nodata\\
$1253-055$ & $1.36$ & $-126.0$ & $1.04$ & $24.91$ & $-0.10\pm1.65$ & $<0.32$ & \nodata\\
$1253-055$ & $2.83$ & $-122.0$ & $0.52$ & $38.68$ & $+0.38\pm1.47$ & $<0.56$ & \nodata\\
$1253-055$ & $4.99$ & $-122.7$ & $0.54$ & $10.93$ & $-1.40\pm1.47$ & $<0.55$ & \nodata\\
$1548+056$ & $0.63$ & $-18.4$ & $1.66$ & $3.24$ & $-0.38\pm1.68$ & $<0.20$ & \nodata\\
$1641+399$ & $1.10$ & $-90.0$ & $1.53$ & $4.14$ & $+1.05\pm1.66$ & $<0.22$ & \nodata\\
$1641+399$ & $2.31$ & $-95.0$ & $0.33$ & $3.43$ & $-0.77\pm0.96$ & $<0.58$ & \nodata\\
$1928+738$ & $0.98$ & $156.0$ & $0.52$ & $1.58$ & $+1.09\pm0.61$ & $<0.33$ & \nodata\\
$2128-123$ & $1.12$ & $-153.4$ & $1.78$ & $1.72$ & $+3.86\pm1.63$ & $+0.22\pm0.09$ & 2.4\\
$2131-021$ & $0.76$ & $113.2$ & $0.61$ & $1.92$ & $+0.46\pm0.63$ & $<0.21$ & \nodata\\
$2134+004$ & $0.80$ & $-90.0$ & $0.77$ & $8.54$ & $+2.21\pm0.94$ & $+0.29\pm0.12$ & 2.4\\
$2134+004$ & $2.02$ & $-98.5$ & $1.33$ & $4.99$ & $+2.99\pm1.56$ & $<0.34$ & \nodata\\
$2145+067$ & $0.71$ & $135.0$ & $1.83$ & $3.01$ & $+0.59\pm1.76$ & $<0.19$ & \nodata\\
$2223-052$ & $0.85$ & $45.0$ & $0.63$ & $6.03$ & $+1.20\pm0.76$ & $<0.31$ & \nodata\\
$2230+114$ & $1.20$ & $138.4$ & $0.32$ & $7.05$ & $+0.11\pm0.45$ & $<0.28$ & \nodata\\
$2243-123$ & $1.60$ & $-3.6$ & $0.43$ & $2.30$ & $+0.60\pm0.57$ & $<0.27$ & \nodata\\
$2251+158$ & $0.80$ & $-90.0$ & $3.11$ & $1.10$ & $+11.94\pm3.20$ & $+0.38\pm0.10$ & 3.7\\
\enddata
\tablecomments{Columns are as follows: (1) IAU Name (B1950); (2) Radial position of the component relative to the jet core;  
(3) Structural position angle of the component relative to the jet core;
(4) Peak Stokes $I$ of the component; 
(5) Fractional linear polarization of the component; (6) Maximum circular polarization measured within two pixels of
the component peak; (7) Fractional circular polarization corresponding to Stokes-$V$ in column (6);
(8) Significance of the circular polarization measurement:  $|m_c|/\Delta m_c$.}
\end{deluxetable}

\begin{deluxetable}{lc}
\tablecolumns{2}
\tablewidth{0pc}
\tabletypesize{\scriptsize}
\tablenum{4}
\tablecaption{Kendall-Tau Test for Correlations with Fractional Core CP.\label{t:tau}}
\tablehead{\colhead{Property} & \colhead{$p$}}
\startdata
Apparent Optical V Magnitude            & $0.07$\\
Optical V Magnitude                     & $0.78$\\
Redshift                                & $0.33$\\
Galactic Latitude, $|b|$                & $0.28$\\
Luminosity Distance                     & $0.33$\\
Angular Size Distance                   & $0.10$\\
Maximum Jet Speed, $\beta_{app}$        & $0.68$\\
Total 15 GHz Flux                       & $0.57$\\
Maximum Total 15 GHz Flux (any epoch)   & $0.44$\\
Total 15 GHz Luminosity                 & $0.38$\\
Maximum Total 15 GHz Luminosity (any epoch) & $0.66$\\
15 GHz Flux of Core                     & $0.94$\\
15 GHz Luminosity of Core               & $0.30$\\
15 GHz Flux of Jet                      & $0.06$\\
15 GHz Luminosity of Jet                & $0.92$\\
Core to Jet Flux Ratio                  & $0.05$\\
Core to Total Flux Ratio                & $0.04$\\
Total Fractional Linear Polarization    & $0.44$\\
Fractional Linear Polarization of Jet   & $0.46$\\
Fractional Linear Polarization of Core  & $0.19$\\
\enddata
\tablecomments{The $p$ value indicates the probability (1 = 100\%) that
no correlation exists between the (absolute) fractional circular polarization
of the core, $|m_c|$, and the listed property.  References for optical
data are given in \citet{LH05}.  Proper motion data is from \citet{K04}. 
Radio data is from \citet{LH05}.}
\end{deluxetable}

\begin{deluxetable}{lcc}
\tablecolumns{3}
\tablewidth{0pc}
\tabletypesize{\scriptsize}
\tablenum{5}
\tablecaption{Comparison of Core CP at 5 GHz and 15 GHz.\label{t:ba18}}
\tablehead{\colhead{IAU Name} & \colhead{5 GHz $m_c$ (\%)} & \colhead{15 GHz $m_c$ (\%)}}
\startdata
0215$+$015 & $< 0.21$ & $+0.28\pm0.12$ \\
0336$-$019 & $+0.19\pm0.06$ & $< 0.18$ \\
0403$-$132 & $< 0.16$ & $< 0.18$ \\
0420$-$014 & $< 0.11$ & $< 0.16$ \\
0605$-$085 & $< 0.10$ & $< 0.26$ \\
0607$-$157 & $-0.18\pm0.05$ & $< 0.21$ \\
0851$+$202 & $< 0.10$ & $-0.20\pm0.08$ \\
0906$+$015 & $< 0.43$ & $< 0.19$ \\
1055$+$018 & $< 0.13$ & $+0.32\pm0.09$ \\
1156$+$295 & $< 0.11$ & $-0.27\pm0.09$ \\
1253$-$055 & $< 0.11$ & $+0.30\pm0.08$ \\
1334$-$127 & $-0.12\pm0.05$ & $+0.29\pm0.10$\\
1413$+$135 & $< 0.18$ & $< 0.34$ \\
1502$+$106 & $< 0.15$ & $< 0.20$ \\
1510$-$089 & $< 0.24$ & $+0.20\pm0.09$ \\
1546$+$027 & $-0.25\pm0.05$ & $< 0.24$ \\
2201$+$171 & $+0.24\pm0.06$ & $+0.19\pm0.09$\\
2223$-$052 & $+0.18\pm0.04$ & $< 0.22$ \\
2227$-$088 & $< 0.12$ & $< 0.18$ \\
2230$+$114 & $-0.09\pm0.04$ & $< 0.19$ \\
2243$-$123 & $-0.14\pm0.04$ & $< 0.25$ \\
2251$+$158 & $+0.08\pm0.04$ & $+0.23\pm0.10$\\
2345$-$167 & $< 0.11$ & $+0.40\pm0.11$\\
\enddata
\tablecomments{The 15 GHz data are from Tables 1 and 2 of this paper. The 5 GHz data 
(epoch 1996 December 15) are from \citet{HAW01}}
\end{deluxetable}


\clearpage

\begin{thebibliography}{dummy}
\expandafter\ifx\csname natexlab\endcsname\relax\def\natexlab#1{#1}\fi

\bibitem[Alberdi et al.(1993)]{AMZ93} Alberdi, A., Marcaide, 
J.~M., Marscher, A.~P., Zhang, Y.~F., Elosegui, P., Gomez, J.~L., \& 
Shaffer, D.~B.\ 1993, \apj, 402, 160 
 
\bibitem[Aller et al.(2003)]{AAP03} Aller, H.~D., Aller, 
M.~F., \& Plotkin, R.~M.\ 2003, \apss, 288, 17 

\bibitem[{{Beckert} \& {Falcke}(2002)}]{BF02}
{Beckert}, T. \& {Falcke}, H. 2002, \aap, 388, 1106

\bibitem[{{Bower} {et~al.}(1999){Bower}, {Falcke}, \& {Backer}}]{BFB99}
{Bower}, G.~C., {Falcke}, H., \& {Backer}, D.~C. 1999, \apjl, 523, L29

\bibitem[{{Bower} {et~al.}(2002){Bower}, {Falcke}, \& {Mellon}}]{BFM02}
{Bower}, G.~C., {Falcke}, H., \& {Mellon}, R.~R. 2002, \apjl, 578, L103

\bibitem[{{Brunthaler} {et~al.}(2001)}]{BBFM01}
{Brunthaler}, A., {Bower}, G.~C., {Falcke}, H., \& {Mellon}, R.~R. 2001, \apjl,
  560, L123

\bibitem[Celotti \& Fabian(1993)]{CF93} Celotti, A., \& 
Fabian, A.~C.\ 1993, \mnras, 264, 228 

\bibitem[Dhawan et al.(1998)]{DKR98} Dhawan, V., Kellerman, 
K.~I., \& Romney, J.~D.\ 1998, \apjl, 498, L111 

\bibitem[{{Fender} {et~al.}(2000){Fender}, {Rayner}, {Norris}, {Sault}, \&
  {Pooley}}]{Fend00}
{Fender}, R., {Rayner}, D., {Norris}, R., {Sault}, R.~J., \& {Pooley}, G. 2000,
  \apjl, 530, L29

\bibitem[{{Fender} {et~al.}(2002){Fender}, {Rayner}, {McCormick}, {Muxlow},
  {Pooley}, {Sault}, \& {Spencer}}]{Fend02}
{Fender}, R.~P., {Rayner}, D., {McCormick}, D.~G., {Muxlow}, T.~W.~B.,
  {Pooley}, G.~G., {Sault}, R.~J., \& {Spencer}, R.~E. 2002, \mnras, 336, 39

\bibitem[{{Homan} {et~al.}(2001){Homan}, {Attridge}, \& {Wardle}}]{HAW01}
{Homan}, D.~C., {Attridge}, J.~M., \& {Wardle}, J.~F.~C. 2001, \apj, 556, 113

\bibitem[{{Homan} \& {Wardle}(1999)}]{HW99}
{Homan}, D.~C. \& {Wardle}, J.~F.~C. 1999, \aj, 118, 1942

\bibitem[Homan \& Wardle(2003)]{HW03} Homan, D.~C., \& 
Wardle, J.~F.~C.\ 2003, \apss, 288, 29 

\bibitem[Homan \& Wardle(2004)]{HW04} Homan, D.~C., \& 
Wardle, J.~F.~C.\ 2004, \apjl, 602, L13 

\bibitem[{{Jones}(1988)}]{J88}
{Jones}, T.~W. 1988, \apj, 332, 678

\bibitem[{{Jones} \& {O'Dell}(1977)}]{JOD77}
{Jones}, T.~W. \& {O'Dell}, S.~L. 1977, \apj, 214, 522

\bibitem[Kellermann et al.(2004)]{K04} Kellermann, K.~I., 
et al.\ 2004, \apj, 609, 539 

\bibitem[Komesaroff et al.(1984)]{K84} Komesaroff, M.~M., 
Roberts, J.~A., Milne, D.~K., Rayner, P.~T., \& Cooke, D.~J.\ 1984, \mnras, 
208, 409 

\bibitem[Lavalley et al.(1992)]{LIF92} Lavalley, M., Isobe, 
T., \& Feigelson, E.\ 1992, ASP Conf.~Ser.~ 25: Astronomical Data Analysis 
Software and Systems I, 25, 245 

\bibitem[Lister \& Homan(2005)]{LH05} Lister, M.~L., \&
Homan, D.~C.\ 2005, \aj, in press, astro-ph/0503152 

\bibitem[{{Macquart} {et~al.}(2000){Macquart}, {Kedziora-Chudczer}, {Rayner},
  \& {Jauncey}}]{MKRJ00}
{Macquart}, J.-P., {Kedziora-Chudczer}, L., {Rayner}, D.~P., \& {Jauncey},
  D.~L. 2000, \apj, 538, 623

\bibitem[Marscher(1987)]{M87} Marscher, A.~P.\ 1987, In 
Superluminal Radio Sources, eds. J.~A. Zensus \& T.~J. Pearson 
(Cambridge: Cambridge Univ. Press), 280 

\bibitem[Pacholczyk \& Swihart(1971)]{PS71} 
Pacholczyk, A.~G.~\& Swihart, T.~L.\ 1971, \apj, 170, 405

\bibitem[{{Rayner} {et~al.}(2000){Rayner}, {Norris}, \& {Sault}}]{RNS00}
{Rayner}, D.~P., {Norris}, R.~P., \& {Sault}, R.~J. 2000, \mnras, 319, 484

\bibitem[{{Ruszkowski} \& {Begelman}(2002)}]{RB02}
{Ruszkowski}, M. \& {Begelman}, M.~C. 2002, \apj, 573, 485

\bibitem[{{Sault} \& {Macquart}(1999)}]{SM99}
{Sault}, R.~J. \& {Macquart}, J.-P. 1999, \apjl, 526, L85

\bibitem[Shepherd(1997)]{S97} Shepherd, M.~C.\ 1997, ASP 
Conf.~Ser.~125: Astronomical Data Analysis Software and Systems VI, 125, 77 


\bibitem[V{\' e}ron-Cetty \& V{\' e}ron(2003)]{VVC03} V{\' 
e}ron-Cetty, M.-P., \& V{\' e}ron, P.\ 2003, \aap, 412, 399 

\bibitem[Walker et al.(1994)Walker, Romney, \& Benson]{WRB94} Walker, 
R.~C., Romney, J.~D., \& Benson, J.~M.\ 1994, \apjl, 430, L45

\bibitem[{{Wardle}(1971)}]{W71}
{Wardle}, J.~F.~C. 1971, \aplett, 8, 183

\bibitem[Wardle \& Homan(2003)]{WH03} Wardle, J.~F.~C., \& 
Homan, D.~C.\ 2003, \apss, 288, 143 

\bibitem[{{Wardle} {et~al.}(1998)}]{WHOR98}
{Wardle}, J.~F.~C., {Homan}, D.~C., {Ojha}, R., \& {Roberts}, D.~H. 1998, \nat,
  395, 457

\bibitem[{{Weiler} \& {de Pater}(1983)}]{WdP83}
{Weiler}, K.~W. \& {de Pater}, I. 1983, \apjs, 52, 293

\bibitem[Zavala \& Taylor(2002)]{ZT02} Zavala, R.~T., \& 
Taylor, G.~B.\ 2002, \apjl, 566, L9 

\end{thebibliography}
\end{document}